# Statistical Properties of Cold Streams in Massive Star-forming Halos in TNG50

Isabel Medlock[1], Daisuke Nagai[1,2], Nir Mandelker[3], Volker Springel[4], Frank C. van den Bosch[1,2], Elad Zinger[3], and Barry T. Chiang[1]

[1] Department of Astronomy, Yale University, New Haven, CT 06520, USA; isabel.medlock@yale.edu
[2] Department of Physics, Yale University, New Haven, CT 06520, USA
[3] Centre for Astrophysics and Planetary Science, Racah Institute of Physics, The Hebrew University, Jerusalem 91904, Israel
[4] Max-Planck-Institut für Astrophysik, Karl-Schwarzschild-Str. 1, 85748 Garching, Germany
*Received 2025 November 26; revised 2026 February 24; accepted 2026 February 27; published 2026 March 25*

## Abstract

Cold, dense streams of gas are predicted to penetrate deeply into massive ($\gtrsim 10^{12}\,M_\odot$) halos at cosmic noon ($z \sim 4$–2), fueling galaxies to sustain high star formation rates. We investigate the prevalence of such cold streams in TNG50 over the range $z = 4$–0, using a novel algorithm to automatically detect cold streams in simulated halos. We qualitatively and quantitatively characterize the geometric and physical properties of the detected streams over cosmic time. We find that cold streams are ubiquitous in massive halos at cosmic noon, occurring in >80% of such systems down to $z = 1$, before becoming rare by $z = 0$. At their peak prevalence ($z = 2$–1), streams are often found in roughly coplanar, three-stream configurations. These streams generally exhibit a dense and cool core, surrounded by a diffuse and warmer envelope. However, we find that, in TNG50, these streams typically disrupt in the outer halo and do not penetrate efficiently to the central galaxy, with the total mass inflow from streams peaking at $z = 2$. Our results underscore the importance of cold streams in fueling galaxies at early times, but they highlight the need for higher-resolution simulations to fully capture their survival and impact at later epochs. Future cosmological zoom-in simulations, with better resolution in the CGM, will be essential to resolve turbulent mixing layers and feedback–inflow interactions that determine whether cold streams can reach the galactic disk.

## 1. Introduction

Gas accretion onto galaxies is a central driver of galaxy evolution, supplying the fuel for star formation, setting the angular momentum and size of disks, and driving turbulence and instabilities. In the classical picture, gas is accreted in two modes: a hot mode in which gas falls into a halo quasi-spherically and is shock-heated near the virial radius $R_{\rm vir}$, and a cold mode in which gas flows along streams that penetrate into the halo. In general, the cold mode dominates at high redshifts and low-mass halos, while the hot mode dominates at low redshifts and high-mass halos (Y. Birnboim & A. Dekel 2003; D. Kereš et al. 2005; A. Dekel & Y. Birnboim 2006).

However, this simple picture breaks down during cosmic noon ($z \sim 2$–4), the peak of cosmic star formation and structure growth (P. Madau & M. Dickinson 2014). During this epoch, massive halos with baryonic mass $\sim 10^{11} M_\odot$ are observed to host rapidly star-forming, rotationally supported disks with star formation rates (SFRs) of $\sim 100\,M_\odot$ yr$^{-1}$ (e.g., R. Genzel et al. 2006, 2008; N. M. Schreiber et al. 2006; D. M. Elmegreen et al. 2007; D. P. Stark et al. 2008; N. M. Förster Schreiber et al. 2009; E. Wisnioski et al. 2015), significantly higher than the estimated current SFR of a few $M_\odot$/yr of the Milky Way (e.g., T. P. Robitaille & B. A. Whitney 2010; L. Chomiuk & M. S. Povich 2011; D. Elia et al. 2022; E. Zari et al. 2023). To sustain such high SFRs, the rate of gas accretion onto the galaxy must closely follow the cosmological accretion rate onto the halo, implying that a majority of this gas must efficiently make its way to the central galaxy.

Indeed, there is a critical regime at roughly $z \geqslant 2$ for halos with $M_h > 10^{12} M_\odot$, where massive hot halos residing at the nodes of cosmic filaments accrete gas through streams of cold, dense gas flowing along the filaments, penetrating deep into the hot halo (D. Kereš et al. 2005; A. Dekel & Y. Birnboim 2006; A. Dekel et al. 2009b; D. Kereš et al. 2009; F. van de Voort et al. 2011b, 2012). Due to high densities ($n \gtrsim 10^{-2}$ cm$^{-3}$) and short cooling times, these cold streams can penetrate the hot halo without being shocked. These streams deliver not only mass but also angular momentum, shaping the structure of the galactic disk (T. Kimm et al. 2011; C. Pichon et al. 2011; K. R. Stewart et al. 2013, 2017; M. Danovich et al. 2015). They are also believed to trigger violent disk instabilities (VDI), leading to the formation of giant star-forming clumps and dramatic compaction events (A. Dekel et al. 2009a; A. Dekel & A. Burkert 2014; A. Zolotov et al. 2015; S. Inoue et al. 2016; S. Tacchella et al. 2016a, 2016b; O. Ginzburg et al. 2025; N. Mandelker et al. 2025).

Although cold streams are difficult to detect observationally, studies of the circumgalactic medium (CGM) around massive high-$z$ galaxies in both absorption (e.g., M. Fumagalli et al. 2011; T. Goerdt et al. 2012; F. van de Voort et al. 2012; N. Bouché et al. 2013, 2016; J. X. Prochaska et al. 2014) and emission (e.g., C. C. Steidel et al. 2000; Y. Matsuda et al. 2006, 2011; D. C. Martin et al. 2014; M. Fumagalli et al. 2017; D. Martin et al. 2019; H. Umehata et al. 2019; E. Daddi et al. 2021) have detected large amounts of cold gas with spatial and kinematic properties consistent with the predictions of cold streams. In cosmological simulations, cold streams are ubiquitous and supply halos with gas at rates comparable to the predicted cosmological accretion rate and the observed SFR, suggesting that a significant fraction of this stream gas must reach the central galaxy (e.g., D. Kereš et al. 2005; P. Ocvirk et al. 2008; A. Dekel et al. 2009b; D. Ceverino et al. 2010; F. van de Voort et al. 2011b). Despite their prominence in simulations, the occurrence and properties of cold streams have not been fully quantified across redshift. Many prior studies focus on tracking accretion rates, cold-mode fractions, or analyzing small halo samples (e.g., F. van de Voort et al. 2011b; D. Nelson et al. 2016; J. Suresh et al. 2019; S. Waterval et al. 2025). Some work has been done to quantify the number of streams present in halos over cosmic time in suites of zoom-in simulations (R. Cen 2014; S. Waterval et al. 2025), but not in large-volume cosmological boxes. Additionally, there are varying depths of penetration and stream morphologies predicted by different simulation codes (e.g., D. Kereš et al. 2005; D. Ceverino et al. 2010; C.-A. Faucher-Giguère et al. 2010; D. Nelson et al. 2013; M. Danovich et al. 2015).

To enable robust comparisons between simulations and establish a clear benchmark, a standardized quantitative approach is needed. In this work, we develop and apply such a method and investigate the statistical occurrence and properties of cold streams around massive galaxies from $z = 4$–$0$ in TNG50. We present a novel algorithm for automatically detecting cold streams in simulated halos. We analyze the evolution of stream frequency and properties such as temperature, density, and mass flux over cosmic time, with a focus on cosmic noon.

This paper is structured as follows. In Section 2, we describe the TNG50 simulation that we analyze. In Section 3, we describe our algorithms for detecting cold streams and analyzing their properties. In Section 4, we present the results of our analysis and discuss the implications and limitations of our work in Section 5. Finally, in Section 6, we summarize our findings and provide our outlook for the future.

## 2. Simulations

We conduct this study using the TNG50 simulation, which strikes a balance between providing a statistical sample of halos and achieving the resolution needed to identify penetrating cold streams.

IllustrisTNG is a suite of magneto-hydrodynamical cosmological simulations performed with the AREPO moving-mesh code (V. Springel 2010). TNG50, the highest-resolution run in the suite, evolves a periodic box of side length $L_{\rm box} = 35$ cMpc $h^{-1}$ ($\approx$51.7 cMpc) with $2160^3$ resolution elements for each of gas and dark matter, respectively, at initialization. The simulations start at $z = 127$ and evolve to $z = 0$, solving the coupled dynamics of dark matter, baryons, stars, and black holes while incorporating a comprehensive galaxy formation model described in R. Weinberger et al. (2017) and A. Pillepich et al. (2018a).

The IllustrisTNG simulations are run with cosmologically motivated initial conditions created using the Zeldovich approximation and the N-GenIC code (V. Springel 2015). The cosmology is consistent with the Planck 2015 results: $\Omega_\Lambda = 0.6911$, $\Omega_m = 0.3089$, $\Omega_b = 0.0486$, $\sigma_8 = 0.8159$, $n_s = 0.9667$, and $h = 0.6774$.

The target baryon mass is $m_{\rm baryon} = 5.7 \times 10^4 M_\odot\,{\rm h}^{-1}$, and the mass of dark matter particles is $m_{\rm DM} = 3.1 \times 10^5 M_\odot\,{\rm h}^{-1}$. The gravitational softening of the collisionless components is $\epsilon_{\rm DM}$, $\star$, wind $= 0.575$ ckpc until $z = 1$, after which it is fixed at 0.288 kpc. The gravitational softening of the gas component is adaptive and proportional to the effective cell radius ($\epsilon_{\rm gas} = 2.5 r_{\rm cell}$) with a minimum of $\epsilon_{\rm gas,min} = 0.074$ ckpc. At $z = 0$, the minimum physical gas cell radius is $r_{\rm cell,min} = 8$ pc, the median gas cell radius is $\bar{r}_{\rm cell} = 5.8$ kpc, and the mean star-forming gas cell radius is $\bar{r}_{\rm cell,SF} = 138$ pc.

Halo catalogs are produced using a standard friends-of-friends (FoF) algorithm applied to the dark matter particles. The subhalo information is computed using the SUBFIND algorithm (V. Springel et al. 2001), modified to include additional baryonic properties.

TNG50 is presented and described in detail in D. Nelson et al. (2019, 2018) and A. Pillepich et al. (2018b), and is publicly available.[5]

## 3. Methods

In this section, we describe how we identify halos with cold streams and analyze the properties of the streams.

### 3.1. Halo Selection through Visual Inspection

We begin by selecting a test sample of candidate halos fed by cold streams at $z = 2$ through visual inspection. We choose all halos with $M_h \gtrsim 10^{12} M_\odot$, as this is the regime at $z \gtrsim 2$ in which we expect "cold in hot" accretion. That is, cold-mode accretion in a shock-heated medium, as opposed to the cold-mode accretion in lower-mass halos that do not exhibit the shock-heated medium (A. Dekel & Y. Birnboim 2006). Using the publicly available TNG Visualize Halos and Galaxies tool,[6] we identify halos that appear to have had no recent mergers (approximately spherical) and feature infalling, dense filamentary structures. This results in an initial sample of 20 halos with masses ranging from $M_h = 10^{11.95} M_\odot$ to $10^{13.27} M_\odot$. The TNG Visualize tool produces maps that only project gas cells associated with the halo as defined by the FoF algorithm and do not show the full extent of the halos' environments. We use this sample to develop and calibrate our analysis pipeline for detecting cold streams, creating projections in Cartesian and spherical coordinates, and identifying the maximum penetration depth of the streams, as described in the following sections.

### 3.2. Cold Stream Finding Algorithm

To develop a consistent and unbiased method for identifying streams beyond visual inspection, we create an algorithm to identify cold stream-like structures for a given halo, which we illustrate in Figure 1. This pipeline processes cell-level data for a halo and its surrounding environment within $3\,R_{200c}$[7] and returns the number of cold streams.

#### 3.2.1. Cold Stream-like Gas Selection

We first select all gas cells between $0.15 < r/R_{200c} < 3$ from the center of the halo to eliminate contamination from the central galaxy. In addition, we filter out cells associated with subhalos to prevent contamination. We identify these gas cells

5 https://www.tng-project.org/data/

6 https://www.tng-project.org/data/vis/

7 $R_{200c}$, defined as the radius of a sphere within which the mean density is 200 times the critical density of the Universe.

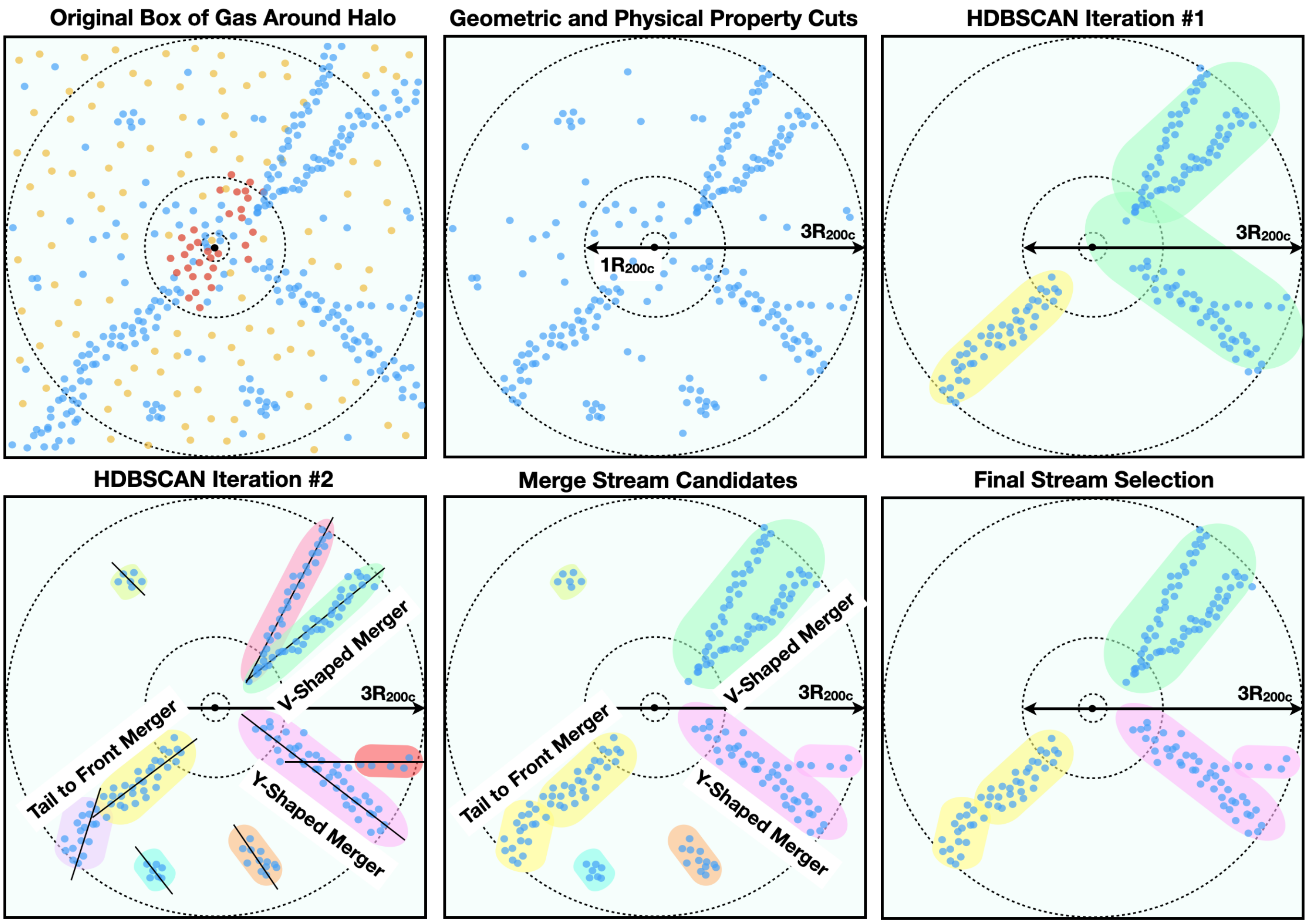

**Figure 1.** Schematic diagram of the stream identification algorithm. In the first panel, we start with all gas cells in a subbox centered around the halo. In the second panel, we select gas within 0.15−3 $R_{200c}$ from the halo center and filter for cold stream-like gas. In the third and fourth panels, we illustrate two examples of the clustering algorithm. In the fifth panel, we demonstrate the merging algorithm. In the sixth panel, we illustrate the final stream selection based on geometric properties of the candidates.

as those within two times the gas half-mass radius of each subhalo, with a minimum radius of 5 kpc.

Next, we select "cold stream-like" gas cells by applying the following cuts on velocity, temperature, and density:

1. To ensure that the gas is infalling, we institute a maximum radial velocity of the gas of $-0.2 \times V_{\rm vir}$. Additionally, we require that 80% of the total velocity (with respect to the host halo frame) be in the radial direction ($v_{\rm rad}/v_{\rm tot} > 0.8$), to ensure predominantly radial infall.
2. We filter the gas cells to have temperatures of $5 \times 10^3\,{\rm K} \leqslant T \leqslant 2.5 \times 10^5\,{\rm K}$. For star-forming cells, we set the temperature to $T = 10^3$ K, before performing this cut. The lower limit[8] of $T \geqslant 5 \times 10^3$ K mitigates the inclusion of small cold cloudlets, while the upper limit follows previous literature, where $T = 2.5 \times 10^5$ K marks the transition between cold- and hot-mode accretion (e.g., D. Kereš et al. 2005, 2009; F. van de Voort et al. 2011b; C. A. Correa et al. 2018b; S. Waterval et al. 2025).
3. Cold streams are dense, allowing them to survive penetration into hot halos. We require a density of at least $n_H \geqslant 10^{-4}\,{\rm cm}^{-3}$. A maximum density of $n_H \leqslant 1\,{\rm cm}^{-3}$ is set to filter out small cloudlets.

These cuts are physically motivated and have precedent in similar studies (e.g., D. Kereš et al. 2005; P. Ocvirk et al. 2008; D. Kereš et al. 2009; C.-A. Faucher-Giguère et al. 2010; F. van de Voort et al. 2011b; C. A. Correa et al. 2018b; S. Waterval et al. 2025). We additionally perform tests to ensure that our results are robust to variations in these parameters, and find that our results remain stable with reasonable variations.

#### *3.2.2. Clustering with HDSCAN to Find Stream Candidates*

We identify cold stream-like gas structures using the Hierarchical Density-Based Spatial Clustering of Applications with Noise (HDBSCAN) algorithm,[9] a hierarchical generalization of the classic Density-Based Spatial Clustering of Applications with Noise (DBSCAN) method (L. McInnes et al. 2017).

[8] While TNG does not allow radiative cooling below $10^4$K, gas can still reach lower temperatures, due to adiabatic expansion.

[9] https://hdbscan.readthedocs.io/en/latest/index.html

DBSCAN identifies clusters as connected regions of high local point density, separated by regions of low density. For each point in the dataset, a neighborhood is defined by a fixed spatial scale ("epsilon") and the number of neighboring points within this region ("min_samples") determines whether it is part of a dense core. Clusters are then built by linking together neighboring core points and including nearby border points that lie within the neighborhood of any core. points that do not meet these criteria are classified as noise. Because DBSCAN relies solely on local density connectivity, it can naturally identify clusters of arbitrary shape, does not require the number of clusters to be specified in advance, and is robust to small-scale noise and outliers. However, it relies on a single density threshold (which is set as a free parameter) to define clusters, which limits its ability to capture structures that vary in density.

HDBSCAN generalizes this framework by performing DBSCAN over a continuum of density thresholds, constructing a hierarchical tree of how clusters appear, merge, or disappear as the density requirements vary. From this cluster tree, HDBSCAN identifies clusters of a minimum size (set by a free parameter) that are statistically stable across a range of density thresholds. This hierarchical approach allows the algorithm to capture structures with varying densities while reducing its sensitivity to parameter choices. In practice, HDBSCAN identifies clusters as locally overdense regions that remain coherent across density scales. Its hierarchical nature allows it to adapt to variable stream densities within and across halos, making it particularly well-suited for identifying filamentary, multi-scale inflow structures embedded in the diffuse CGM.

For each halo, we feed HDBSCAN all gas cells previously identified as cold stream-like gas, representing them as points with coordinates of the center of mass of each cell. We set the number of neighboring points within this region ("min_samples") adaptively for each halo, starting with an initial value. When this value is set higher, the threshold for the minimum number of points in each candidate increases, resulting in fewer stream candidates being identified. However, if this value is set too high, gas substructures may be incorrectly merged, combining individual streams or clumps into a single stream candidate, as illustrated in the third panel of Figure 1.

To ensure that the stream candidates identified by HDBSCAN correspond to physically distinct and well-defined structures, we perform a simple geometric consistency check on each candidate after clustering. We verify that the cross-sectional extent of each stream is not excessively wide, compared to the halo size. For each candidate, we determine its principal axis using principal component analysis (PCA) and construct bounding cylinders aligned with that axis. We then measure how the candidate's points are distributed within these cylinders. Specifically, we require that at least 99% of the points lie within a cylinder of radius $0.75\,R_{200c}$, and that at least 90% are contained within $0.5\,R_{200c}$. Candidates failing to meet these conditions are rejected, ensuring that the identified streams are compact and spatially coherent rather than diffuse or halo-spanning features.

HDBSCAN is then iteratively performed with decreasing values for the number of neighboring points required ("min_samples"), breaking the gas into more substructures until at least 80% of the stream candidates pass this test. We stop the algorithm if the number of stream candidates exceeds 20 or the change in the number of stream candidates between iterations exceeds 5. This approach is necessary to ensure that the algorithm can adapt to the varying densities and morphologies of streams across different halo masses and redshifts.

#### *3.2.3. Consolidating Stream Candidates*

To ensure that the stream candidates remaining after HDBSCAN are not substructures of a single stream, we merge stream candidates based on the following criteria:

1. *V-shaped Mergers.* If two stream candidates meet at their heads (the endpoints pointing toward the halo center) and have principal component axes within 15°, they are merged into a single structure. The fourth panel of Figure 1 shows a schematic example of this case for the streams on the upper right.
2. *Tail-to-front Mergers.* If streams are slightly fragmented in the radial direction, stream candidates are merged if their principal axes are within 45° of each other, and the tail end of one is within 30 kpc of the front of the other. The fourth panel of Figure 1 shows a schematic example of this case in the lower left corner of the box.
3. *Y-shaped Mergers.* Finally, we consider cases where the head of one stream candidate is close to anywhere along the primary axis of another stream candidate. We allow more leeway here compared to the "V-shaped" case, and these candidates are considered merged if their axes are within 53° and the front of one stream candidate is within 30 kpc of the main axis of the other. The fourth panel of Figure 1 shows a schematic example of this case for the streams in the lower right.

We perform this merging iteratively by testing each pair of stream candidates until no further mergers occur. Various angular thresholds were tested to determine the optimal values for the final analysis. Specifically, for V-shaped mergers, we classify these as a single stream rather than two separate streams only if the streams are in close proximity. For tail-to-front and Y-shaped mergers, we set these angular thresholds to be less stringent, to account for the potential curvature and branching of the streams, as shown in the two examples in the lower half of each panel in Figure 1.

#### *3.2.4. Cold Stream Final Selection*

Subsequently, we analyze the geometry of the stream candidates to identify elongated, stream-like structures. As an initial cut, we select stream candidates with a radial extent $\geqslant 1.5\,R_{200c}$. Additionally, we exclude any stream candidates that are less than 5% of the size of the largest stream candidate. We characterize the three-dimensional shape of each remaining stream candidate using the shape tensor (inertia tensor). To determine how elongated the stream candidate is, we derive the axial ratios of the principal axis relative to the secondary ($a/b$) and tertiary axes ($a/c$) from the shape tensor. Higher axial ratios indicate greater elongation along a single dominant direction. We set axial ratio thresholds of 5 and 10, for $a/b$ and $a/c$ respectively, to select filamentary structures, as illustrated in the sixth panel of Figure 1.

This refined methodology ensures the consistent identification and classification of stream-like structures, enhancing our understanding of their dynamics and interactions within the CGM.

One important note is that this algorithm is designed to identify stream candidates in a consistent manner, but it does not recover all gas cells associated with streams. This can lead to underestimation when using the results to quantify total quantities such as mass and mass flux in streams.

### 3.3. Projections

To create slices and projections of the halos and their environments, we start by mapping the cell data to a uniform 3D grid. We include all cells within $5\,R_{200c}$ of each halo, not just those associated with the halo identified via the FoF algorithm.

To construct the uniform 3D grid, each Voronoi cell is mapped onto a Cartesian grid by identifying the set of grid voxels intersected by the cell. We approximate cells as spheres with an effective radius derived from their volume. The fractional overlap between each cell and the intersected grid voxels is estimated geometrically. Extensive quantities like mass are distributed across the intersected voxels in proportion to their estimated fractional overlap with the cell, with the normalization chosen such that the total deposited value equals the original cell value. Intensive quantities like temperature are averaged over the intersected voxels using overlap-based weights (and optional mass weighting), ensuring a properly weighted mean rather than a direct summation.

Using the uniform 3D grid, we take 2D Cartesian projections of different gas properties, including density, radial velocity, mass flux, temperature, and entropy. To produce the projections using the uniform 3D grid, we specify the desired axis, center plane, and projection depth. We take the mean along each line of sight for intensive quantities, and the sum for extensive quantities.

In addition, we generate Hammer projections of these quantities in spherical coordinates. For the Hammer projections, we use healpy, an implementation of Hierarchical Equal Area isoLatitude Pixelation of a sphere (HEALPix)[10] (K. M. Górski et al. 2005; A. Zonca et al. 2019), to transform to these coordinates and take the projections. This ensures that every pixel on the map covers the same surface area.

### 3.4. Stream Penetration Algorithm

Using Hammer projections, we develop an algorithm to detect the maximum penetration of streams. This also serves as a secondary check on the number of streams identified by our stream finding algorithm. In cases of significant discrepancies between the number of streams identified by the stream finding algorithm and the stream penetration algorithm, we manually review the halo.

We begin by constructing a Hammer projection of the mass inflow through a radial shell at $r = 1.4\,R_{200c}$, referred to as the *outer shell*. We identify regions of strong inflow by applying a percentile-based threshold (`p_low`) that selects pixels with the highest mass influx. For example, selecting the top 5% of pixels isolates the most prominent inflowing regions in the shell map (see the rightmost panels of Figure 4 for an example of this).

[10] https://healpix.sourceforge.io/

**Table 1**
Halo Sample Used for the Analysis in the Paper

| Redshift ($z$) | # Halos | $\log(\langle M_h/M_\odot\rangle)$ | $\langle$SFR$\rangle$ ($M_\odot$ yr$^{-1}$) |
|---|---|---|---|
| 4 | 15 | 12.24 | 189.2 |
| 3 | 45 | 12.31 | 122.6 |
| 2 | 136 | 12.39 | 69.5 |
| 1 | 230 | 12.53 | 35.8 |
| 0.5 | 248 | 12.60 | 21.3 |
| 0 | 230 | 12.75 | 8.8 |

**Note.** For each redshift, we provide the number of halos that met our mass criteria, the mean halo mass, and the mean SFR.

In each subsequent inner shell, a new mask is generated using the corresponding local percentile threshold at that radius. To maintain the continuity of each stream and prevent artificial fragmentation, we restrict how much the threshold can vary between adjacent shells. Specifically, we define both an upper and a lower bound on the local percentile-based threshold, anchored to the flux values in the outer shell.

For each shell mask, HEALPix is used to identify connected components or clumps within the inflow. We retain the largest $N$ clumps (parameter `max_regions`), ensuring a minimum number of pixels (parameter `min_pixel`) to capture the most significant inflows. Streams are tracked iteratively inward by matching labeled inflow regions across shells, using an intersection-over-union threshold of 0.1 to determine overlap. To account for noise and minor shape evolution, we employ pixel dilation on the inflowing clumps, extending the clump boundaries outward by 2 pixels.

This tracking process defines three key stream metrics: penetration depth (the deepest shell where the stream persists), size (angular area on the map), and strength (mass inflow). For the latter two quantities, we track these over all shells and consider the average values, maximal values, and the value at the outer shell. By refining our methodological approach in this manner, we aim to obtain a detailed and accurate representation of inflowing gas streams, their dynamics, and persistence within the CGM.

To ensure robustness, we perform a parameter sweep and consensus selection. We vary three key parameters (flux threshold, max regions, and min pixels) over a range of values, tracking the stream metrics. We match streams across runs to determine how often each stream is detected. Robust streams are those that appear in at least 50% of the parameter variations. For each robust stream, we compute the median, mean, and maximum penetration depth of the collection of penetration depths from the parameter variations, as well as the mean strength and position on the map.

### 3.5. Halo Sample

For our final analysis, we apply these algorithms to all halos with $M_h \geqslant 10^{11.95} M_\odot$ across redshifts $z \in \{0, 0.5, 1, 2, 3, 4\}$. Table 1 presents the number of halos we analyzed for each redshift, along with the mean halo mass and SFR for reference. Consistent with the growth of structure over cosmic time, at later times, we have larger samples of halos, and the mean halo mass increases monotonically. The mean SFR decreases monotonically with time.

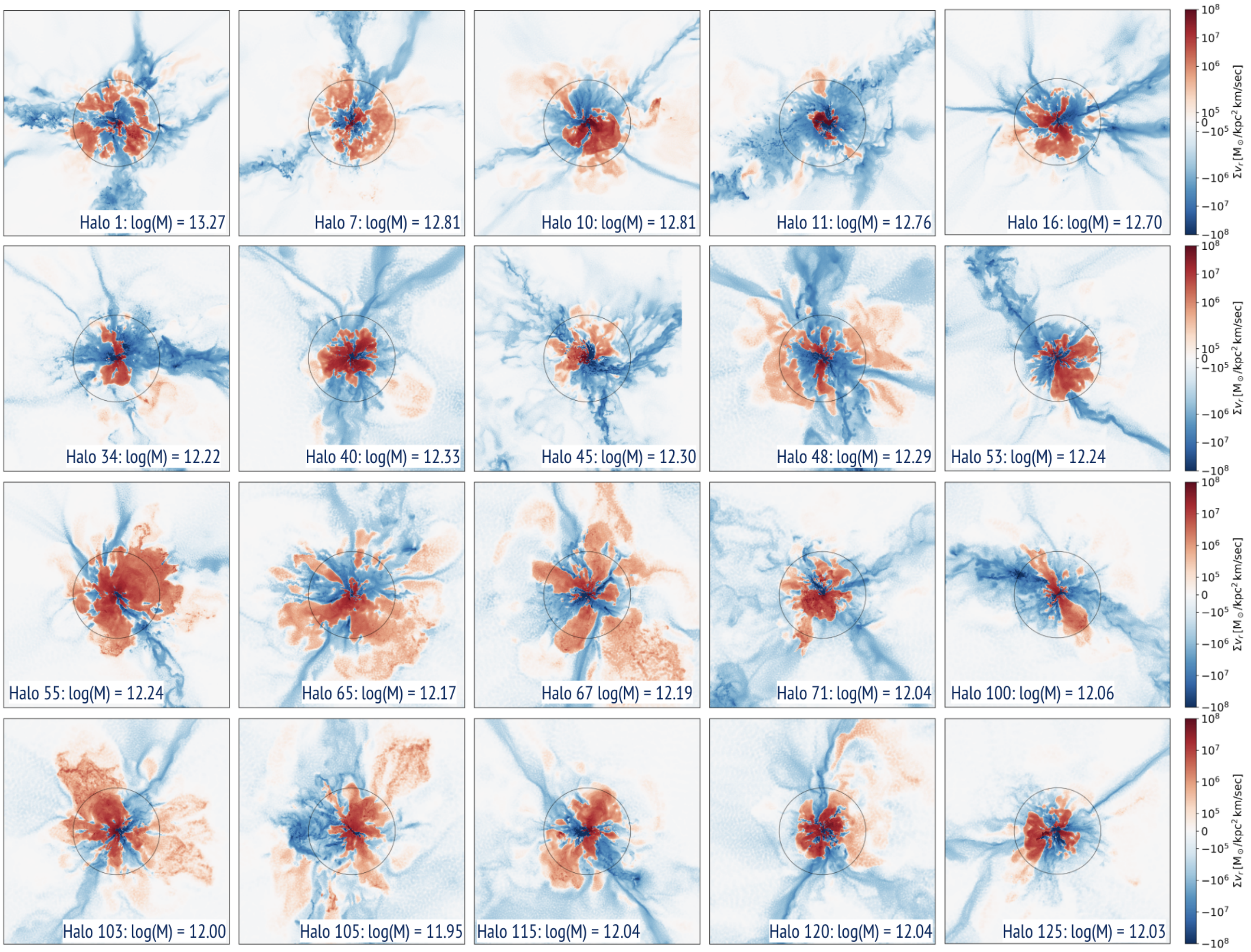

**Figure 2.** A selection of 20 halos with $M_h \geqslant 10^{11.95} M_\odot$ at $z = 2$ in TNG50. For each of the halos, we show a mass flux projection plot, with a depth of 0.1 $R_{200c}$, where blue indicates inflowing mass and red indicates outflowing mass. $R_{200c}$ is demarcated via the dashed black circles for each halo, and each map extends 2.6 $R_{200c}$ on each side to show the broader environment of the halos. At the bottom of each map, we list the halo ID in the simulation as well as the halo mass. These halos demonstrate the prevalence of streams in these types of systems, as well as the diversity in the configurations and properties of the streams.

## 4. Results

In this section, we present the results of our investigation, starting with qualitative results via visual inspection, followed by quantitative statistical results. We relegate an in-depth interpretation of the results and a comparison with previous literature to the discussion section.

### 4.1. Visual Inspection with Halo Maps

We begin with the $z = 2$ snapshot, featuring a sample of 136 massive halos at cosmic noon, the peak of the cosmic star formation rate. Figure 2 showcases a gallery of the 20 halos we initially selected via visual inspection, qualitatively illustrating the prevalence of cold streams during this epoch. For each halo, we present a mass flux projection plot (integrated over a depth of 0.1 $R_{200c}$), where blue denotes inflowing mass and red denotes outflow.

This sample highlights the widespread presence of cold streams in massive halos at $z = 2$ and the variation in their properties and configurations. Although all systems exhibit some form of filamentary inflowing gas structures, each system is distinct. Broadly, there appear to be two types of streams: thicker and more diffuse streams, and thinner and denser ones. Halos with thinner streams, such as Halos 10, 16, and 125, usually have three or more streams feeding the halo. In contrast, halos with thicker streams, such as Halos 11 and 100, often sit between two wide streams on either side. Sometimes, two or more thinner streams merge into a single, thicker stream before entering the halo, as seen in the left-hand streams of Halo 125 and the upper streams of Halo 40. In other cases, such as Halos 34 and 40, there is a combination of thick and thin streams. However, this thick/thin dichotomy is not rigid, as streams of intermediate thickness are also observed in halos such as 7 and 120.

Next, we focus on two illustrative examples to further explore the properties and features of these streams. Figure 3 presents maps at $z = 2$ of Halo 10 (top row) and Halo 120 (bottom row). For each halo, we show the maps of surface density (left column), mass flux (middle column), and temperature weighted by gas cell density (right column) over a projection depth of 0.1 $R_{200c}$. In each map, the inner circle marks $R_{200c}$, while the outer circle marks 2.5 $R_{200c}$.

In the top row, we present Halo 10, characterized by a mass of $M_h = 10^{12.81} M_\odot$. Three prominent streams of cold, dense,

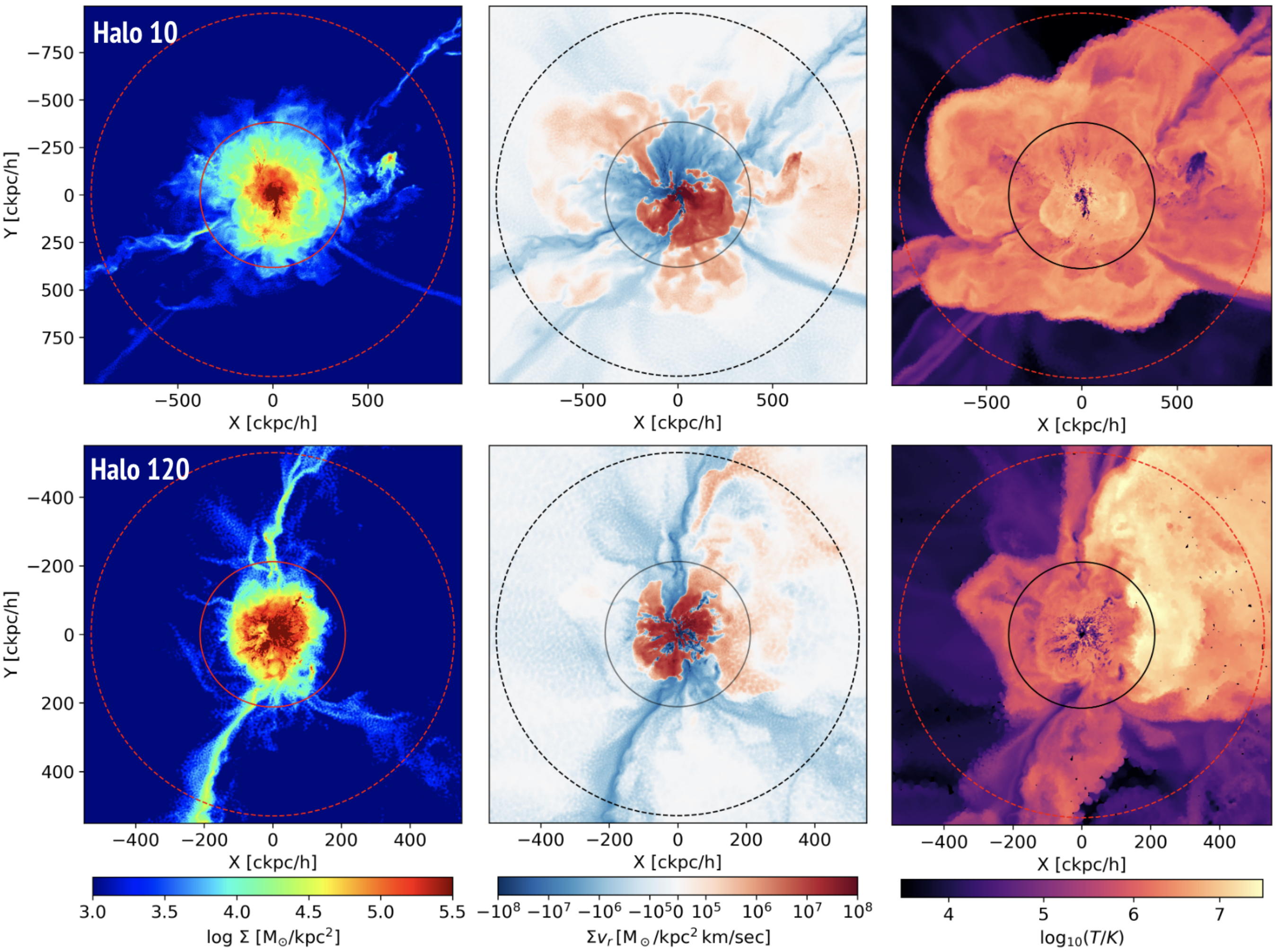

**Figure 3.** Two example halos at $z = 2$ that exhibit cold streams, Halo 10 ($M_h = 10^{12.81} M_{\odot}$; top row) and Halo 120 ($M_h = 10^{12.04} M_{\odot}$; bottom row). We show a surface density map (left column), mass flux map (middle column), and density-weighted temperature map (right column), with a projection depth of 0.1 $R_{200c}$. For each map, $R_{200c}$ is demarcated with the inner circle, and 2.5 $R_{200c}$ with the outer circle. Like in Figure 2, each map extends 2.6 $R_{200c}$ on each side to show the broader environment of the halos. Small fluctuations in the maps are artifacts of sparse sampling, resulting from the thin projection depth and the applied density and mass flux cuts. In the temperature maps, the small black clumps correspond to very cold regions or cells with no data.

inflowing gas are visibly discernible, accompanied by a weaker stream emerging from the lower left corner. The surface density maps reveal that these streams possess a thin, dense core enveloped by a broader, diffuse outer layer. Notably, the dense cores exhibit curves and kinks, deviating from a perfectly straight trajectory.

The mass flux maps display analogous characteristics, with stronger mass inflow concentrated in the dense cores surrounded by a more diffuse envelope. When comparing the inflows with the outflows, we observe that, on large scales, the streams remain relatively thin. Within approximately $R_{200c}$, there is additional inflow that is more isotropic, potentially resulting from recycled outflows. This is corroborated by the disappearance of these isotropic inflows upon the imposition of a low metallicity cut ($<0.1 Z_{\odot}$), whereas the thin streams persist, as illustrated in Figure 13 in the Appendix. In contrast, the outflows are robust within $R_{200c}$, transitioning to a more diffuse, widening structure beyond the halo boundary, such as in Halos 65, 67, and 103. The temperature maps indicate that the streams are significantly colder than the surrounding medium, with the colder cores encased by slightly warmer, diffuse envelopes.

In the bottom row, we present Halo 120, with a mass of $M_h = 10^{12.04} M_{\odot}$. Here, two very prominent streams and a more diffuse third stream are visible. Similar to Halo 10, the prominent streams exhibit thinner, denser cores with noticeable bends, curves, and kinks. The mass-flux projection reveals detailed structures within the streams, including the merging of the two bottom streams within the halo boundaries. In the temperature map, the core of these streams is colder, with temperatures around $10^4$–$10^{4.6}$ K, surrounded by a diffuse, warmer shell.

Figure 4 presents the hammer projections at $r = R_{200c}$, corresponding to the maps in Figure 3. On the left and center columns, the surface density, mass-flux Hammer projections are displayed with a projection depth of 0.1 $R_{200c}$. On the right, we present the mass-flux Hammer projection with a mask to highlight the top 5% of the strongest inflow regions, again with a projection depth of 0.1 $R_{200c}$.

Looking at the entire mass flux projection, we see that high-density areas of cold inflow manifest as streaks rather than self-contained blobs. Closer inspection reveals blob-like regions within these streaks, characterized by higher density in the surface density maps on the left and more pronounced inflow in the mass flux maps in the center. These dense regions of inflow are highlighted in the masked mass-flux Hammer projections, where we can observe several distinct groups of blobs. These correspond to the cold streams seen in the Cartesian maps of Figure 3. These streaks could represent planes or cosmic sheets of inflow in which the streams are embedded (N. Mandelker et al. 2019b; I. Pasha et al. 2023). These projection maps are integral to our stream penetration algorithm, determining the maximum penetration depth of cold streams. In Halo 10, we observe four high-density regions with strong mass inflow. In Halo 120, three peak inflow regions correspond to high-density areas.

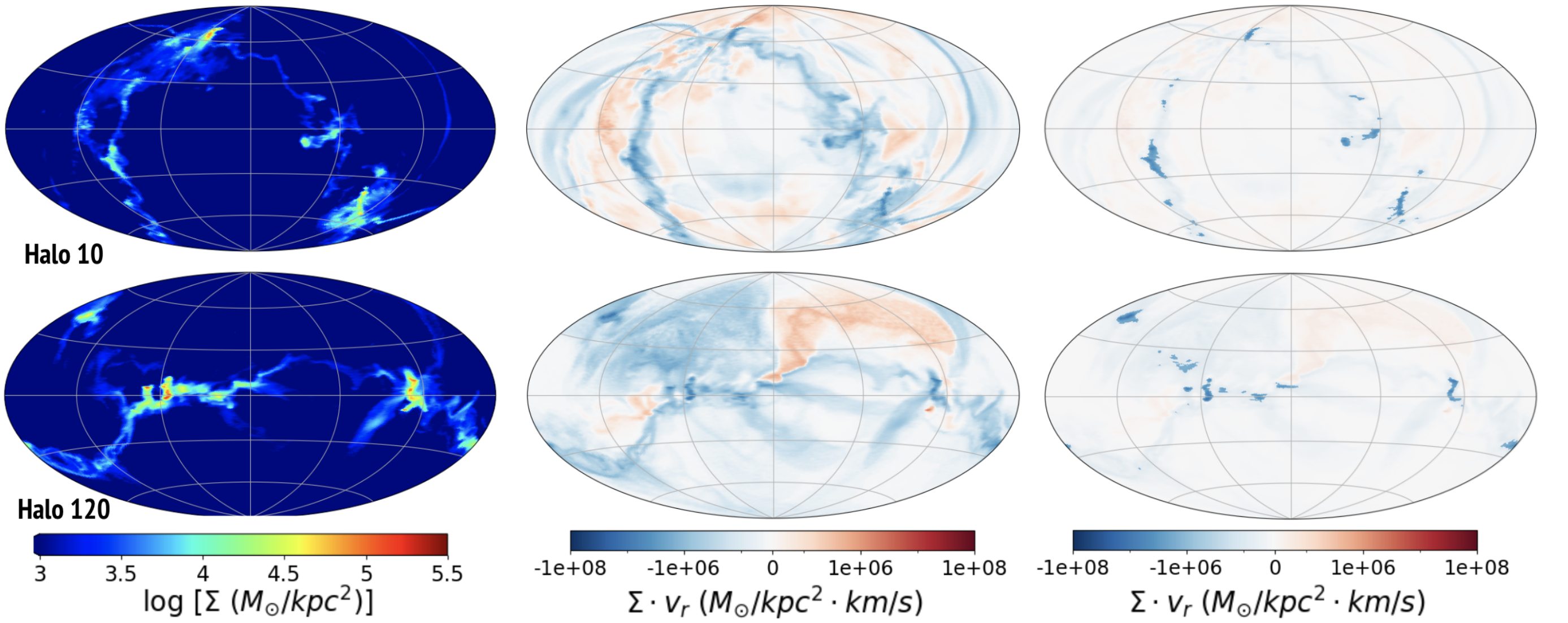

**Figure 4.** Hammer projection maps corresponding to Figure 3. We show a surface density Hammer projection (left column), mass-flux hammer projection (middle column), and the mass-flux Hammer projection masked to highlight strong mass influx (right column), with a projection depth of 0.1 $R_{200c}$ around $r = R_{200c}$.

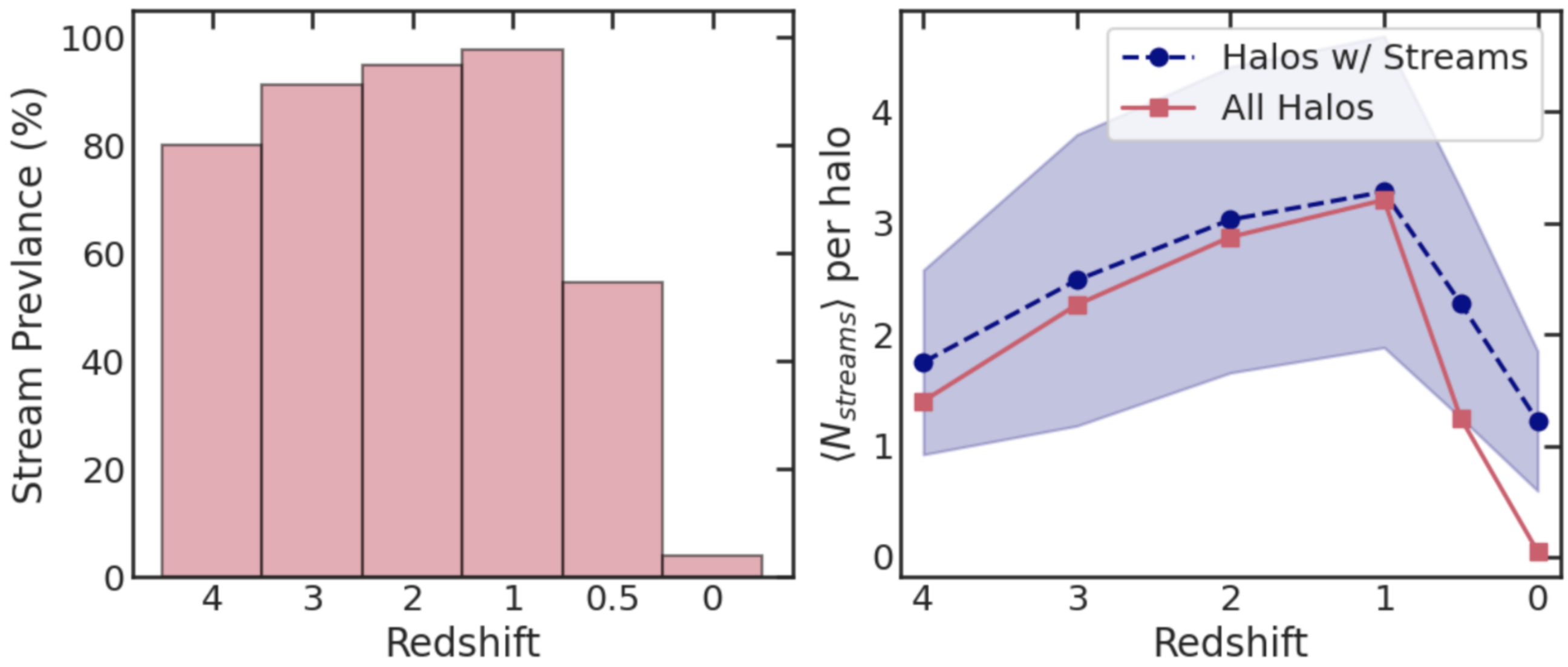

**Figure 5.** Prevalence of streams in massive halos over redshift. Left: percentage of halos with $M_h \geqslant 10^{11.95} M_\odot$ that have at least one stream as a function of redshift. Right: the mean number of streams in a halo, taken over only halos with at least one stream (purple dashed line) and over the entire halo population (pink solid line) as a function of redshift.

Note that the Hammer projections in Figure 4 present a single radial shell at $R_{200c}$. When performing stream identification with these maps, we check for persistent, strong inflows across multiple shells, confirming that these are streams and not infalling small-scale clumps that appear in just one or two shells. By eye, looking at the right-hand panel of Figure 4, more than four or three clumps are apparent; however, these do not persist across multiple shells.

### 4.2. Cold Stream Prevalence in Massive Halos

Next, we conduct a quantitative analysis of the presence of streams in massive halos from $z = 4$ to $z = 0$, using the methods detailed in Section 3. Figure 5 illustrates the prevalence of halos with streams and the number of streams per halo as a function of redshift. It is important to note that, consistent with the growth of cosmic structures over time, the number of massive halos ($M_h \gtrsim 10^{12} M_\odot$) at higher redshifts ($z = 4$–3) is significantly lower (15 and 45 at $z = 4$ and $z = 3$, respectively). This population increases at lower redshifts,

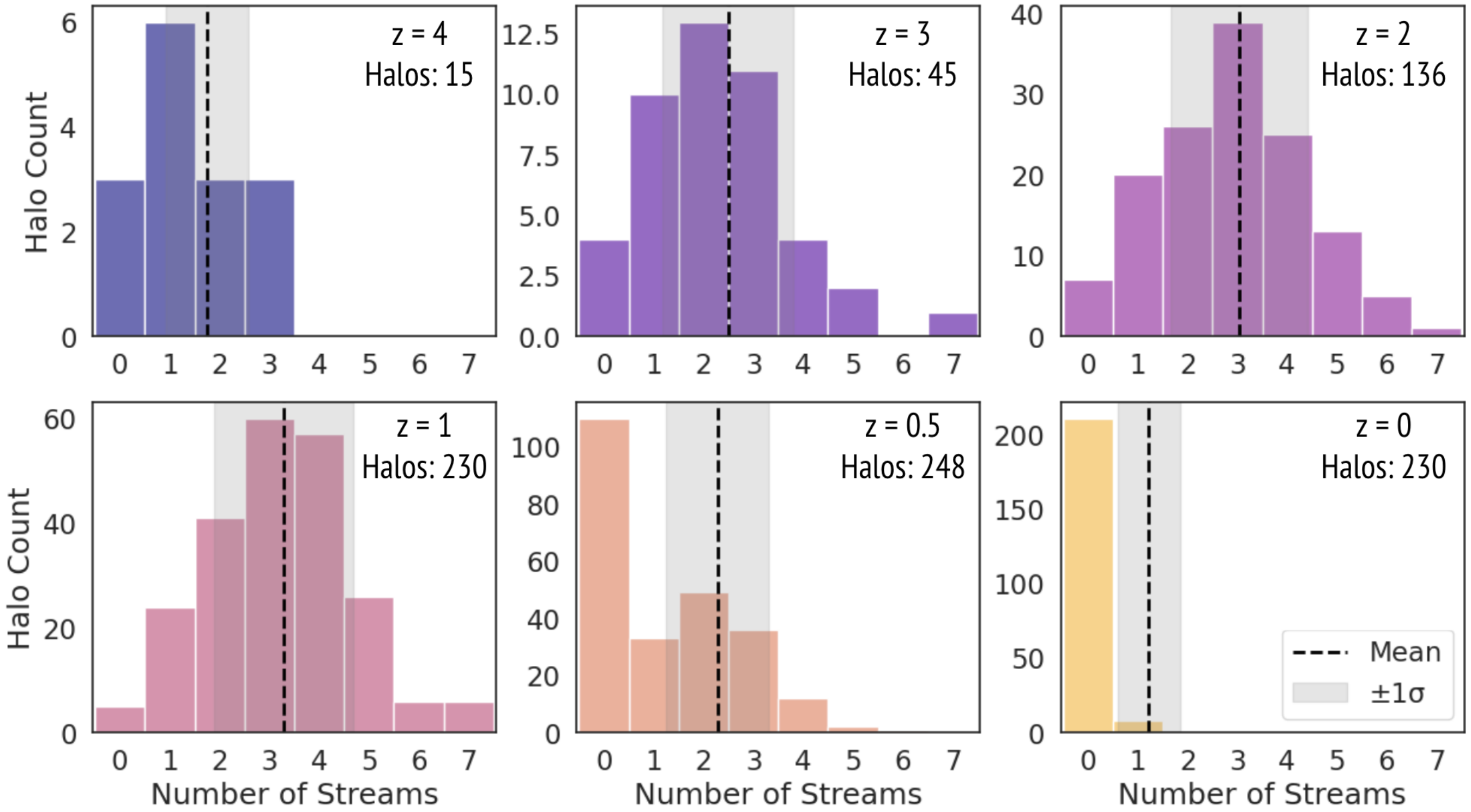

**Figure 6.** The distribution of the number of streams detected in each halo at the indicated redshift from $z = 4$–0. The mean number of streams for halos that have at least one stream is shown with the vertical dashed line, with the standard deviation demarcated by the shaded region.

reaching 136 at $z = 2$, 230 at $z = 1$, and 248 at $z = 0.5$, as shown in Figure 6.

The left panel displays the percentage of halos with $M \geqslant 10^{11.95} M_\odot$ that host at least one stream. Cold streams are prevalent in massive halos from $z = 4$ to $z = 0.5$, but are rare at the present epoch ($z = 0$). From $z = 4$ to $z = 1$, the percentage of halos with streams increases slightly from 80% to 95%. After $z = 1$, the prevalence decreases but remains common (55%) at $z = 0.5$. However, at $z = 0$, there is a drastic decline, and halos with streams are very rare, with only 9 out of 230 halos exhibiting streams.

The right panel of Figure 5 shows the mean number of streams per halo, considering only halos with at least one stream (purple line) and all halos (pink line). At higher redshifts ($z = 4$–1), the difference between these lines is minimal, as most halos possess at least one stream. At lower redshifts, where halos without streams become common, this difference becomes more pronounced. The shaded purple region indicates the standard deviation among halos with at least one stream. The number of streams per halo peaks around $z = 2$–1, with the average configuration featuring three prominent streams, in line with the classic picture of streams (A. Dekel et al. 2009b; M. Danovich et al. 2012). At higher redshifts, the mean number of streams is around two, suggesting that these halos might be embedded in the center of a filament. At lower redshifts, the mean number of streams decreases to around two at $z = 0.5$ and one at $z = 0$.

The six panels of Figure 6 present the full distributions of the number of streams per halo across six redshift snapshots. At $z = 4$, there are 15 massive halos, 12 of which have at least one stream. Half of these halos have only one stream, while the other half have two or three streams. From $z = 3$ to $z = 1$, the number of massive halos increases significantly, with more halos exhibiting three-stream configurations. Some halos even have up to seven detected streams, confirming the diversity seen in Figure 2, where we observed halos with varying stream numbers at $z = 2$. At $z = 1$, four-stream configurations are also common. At $z = 0.5$, the proportion of halos without streams increases sharply, although a significant number still have streams, with two-stream configurations being the most common. Finally, at $z = 0$, only 9 out of 230 halos have at least one stream; eight of these have one stream, and one halo has a three-stream configuration.

### *4.3. Cold Stream Physical Properties*

We now quantitatively calculate the physical properties, including the temperature and density of the cold streams, from the stream cells identified with our stream finding algorithm, as discussed in Section 3. Figure 7 presents the mean temperature, weighted by the density of the gas cells in the detected streams, over cosmic time. The mean over all halos is shown with the dark blue line, with the variance demarcated by the shaded region. The mean of the streams in the 50% comprised of the most-massive host halos is shown with the purple line, and the mean of the streams in the least-massive host halos is shown with the yellow line.

Notably, the mean temperature of the streams increases gradually from $T \sim 10^4$ K at $z = 4$ to $T \sim 10^{4.8}$ K at $z = 0$. The variance in the temperature of the streams increases with cosmic time as well. The black dashed line indicates the classical temperature threshold that defines the transition between cold-mode and hot-mode accretion. As we can see,

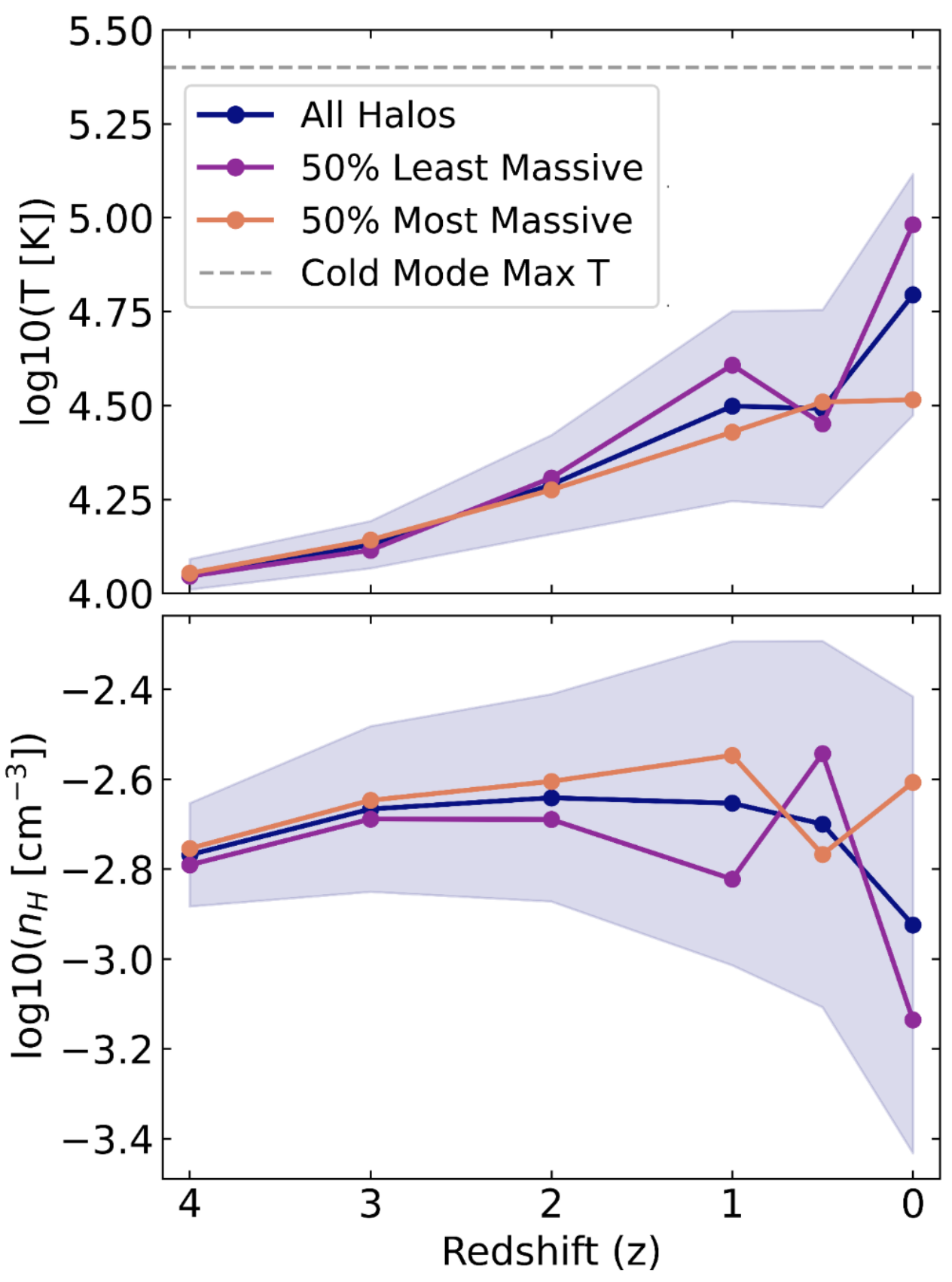

**Figure 7.** Mean physical properties of streams over cosmic time from $z = 4$–$0$. The top panel shows the mean log temperature in Kelvin, while the bottom panel illustrates the mean log hydrogen number density $n_H$ [cm$^{-3}$]. The dark blue line represents the mean values across all halos. Additionally, the mean values are shown for stream host halos binned by the top 50% (most massive) and the bottom 50% (least massive). The shaded regions indicate the standard deviation of the log around the mean of all halos.

the streams detected by the algorithm remain well below this threshold.

In contrast to the increase in temperature over time, the mean density remains relatively constant, hovering between $n_H \sim 10^{-2.8}$–$10^{-2.6}$ cm$^{-3}$. There is a slight drop in density for the few streams at $z = 0$ in the lower-mass halos. The variance in density increases over time. This increase in temperature, paired with a relatively constant density, implies an increase in the streams' thermal pressure over time.

Finally, we examine whether the stream properties exhibit a dependence on host halo mass. At early times ($z = 4$–$2$), there is no apparent dependence of host halo mass on stream temperature. This lack of a trend likely reflects the narrower range of halo masses sampled at these redshifts. At $z = 1$ and $z = 0$, and partly at $z = 2$, we see that more-massive halos host slightly cooler streams. In terms of density, at early times, there is not a significant difference due to host halo mass; however, at later times, we generally see that more-massive halos host denser streams. An exception for both of these cases is at $z = 0.5$, where we see that there is not a significant difference in temperature, but that less-massive halos host denser streams. However, the limited statistics at this redshift preclude a definitive physical interpretation and warrant a future, dedicated study.

### 4.4. *Mass Inflow over Cosmic Time*

A key property that we want to quantify for the streams and for the host halos, in general, is the rate of mass influx over cosmic time. Here, we investigate the global mass accretion rate of our halo sample over time and quantify the contribution of cold streams.

First, in Figure 8, we present the total mass inflow rate (in $M_\odot$/yr) as a function of halo mass, with each halo point colored by the corresponding star formation rate (SFR, $M_\odot$/yr). Superimposed on this is the theoretical baryonic mass accretion rate from A. Dekel et al. (2013). We calculate the mass accretion rate in a thin shell $0.02R_{200c}$ around $R_{200c}$ here and for all the other mass influx calculations.

Our results generally show that, at all redshifts, the mass inflow rate increases with halo mass, accompanied by a rise in the SFR. Regarding redshift evolution, the strength of the mass inflow decreases with lower redshift (considering total mass influx, not just from streams, as shown in Figure 10). The calculated inflow rates are broadly consistent with the predictions from A. Dekel et al. (2013).

Next, in Figure 9, we present the inflow rate over time, broken down into the contributions from specific phases of gas. In the top panel, we show the mean halo mass accretion rate over cosmic time for all inflowing gas in blue, for low-metallicity inflowing gas ($Z < 0.1Z_\odot$) in purple, and cold, low-metallicity inflowing gas ($T < 10^{5.4}$ K, $Z < 0.1Z_\odot$) in yellow. In the bottom panel, we show the fractional ratio of mass accretion from streams to each of these accretion rates, using the same colors. Stream flux is calculated in the same way as the total flux over a given shell, only considering the gas cells that are identified as belonging to streams by the stream finding algorithm described in Section 3.

In the top panel, we see that, in general, the mass accretion rate decreases as a function of cosmic time, in agreement with the results in Figure 8 and A. Dekel et al. (2013). As expected, when we impose cuts on the gas, the mass accretion rate decreases. Indeed, the cold, low-metallicity accretion rate is only ∼20% of the total mass accretion rate and half of the low-metallicity accretion rate. The contribution from streams to this accretion rate is a small fraction of the accretion rate of even the cold, low-metallicity gas. One interesting feature is that, at $z = 2$, there is a strong peak in the fractional contribution of stream accretion to the total cold, low-metallicity mass accretion rate.

Here, it is important to note that, while our algorithm provides a reliable method to identify stream structures, it does not attempt to quantify all gas cells associated with each stream. In particular, the algorithm is tuned to identify the densest core of each stream and does not capture the more diffuse envelope. For the purpose of calculating the average mean and density of the streams, this is fine; however, for computing the mass flux of the streams, where each cell contributes, this is a significant underestimation. For these purposes, then, the analysis serves more as a comparative exercise across redshifts using a consistent method rather than as an absolute quantification of the mass accretion rate derived from streams.

Finally, Figure 10 presents the mass influx from our detected cold streams. In the top panel, we show the mean mass accretion rate at a shell at $R_{200c}$ of individual streams across cosmic time. In the bottom panel, we show the total accretion summed over all streams from all halos at each redshift. This redshift dependence reflects the redshift dependence of the number of massive halos, the average number of streams per halo, and the strength of each individual stream.

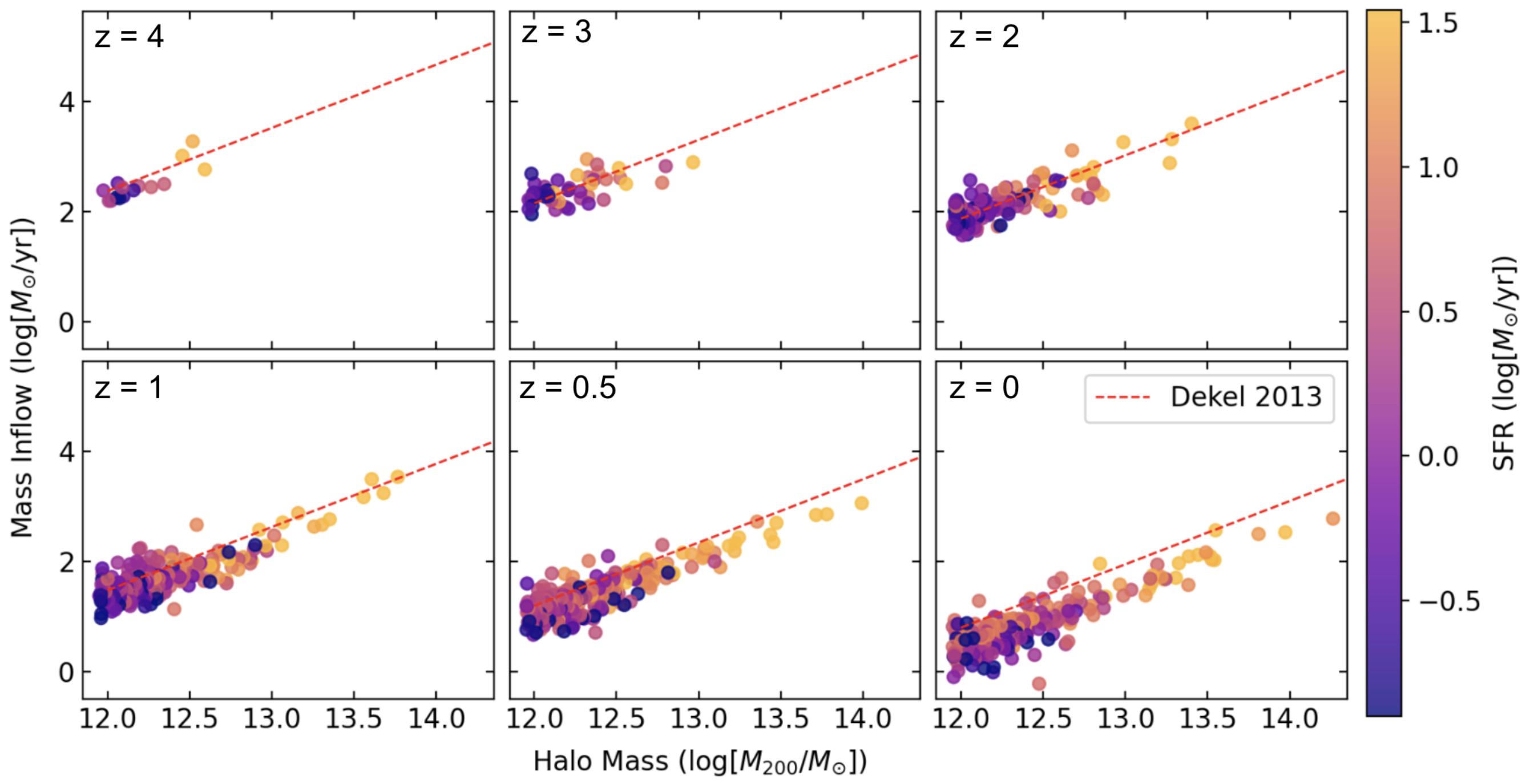

**Figure 8.** The total mass influx ($M_\odot$/yr) at $R_{200c}$ as a function of halo mass, colored by star formation rate at the six redshift snapshots. The dashed red line indicates the theoretical baryonic accretion rate from A. Dekel et al. (2013), as a function of halo mass and redshift.

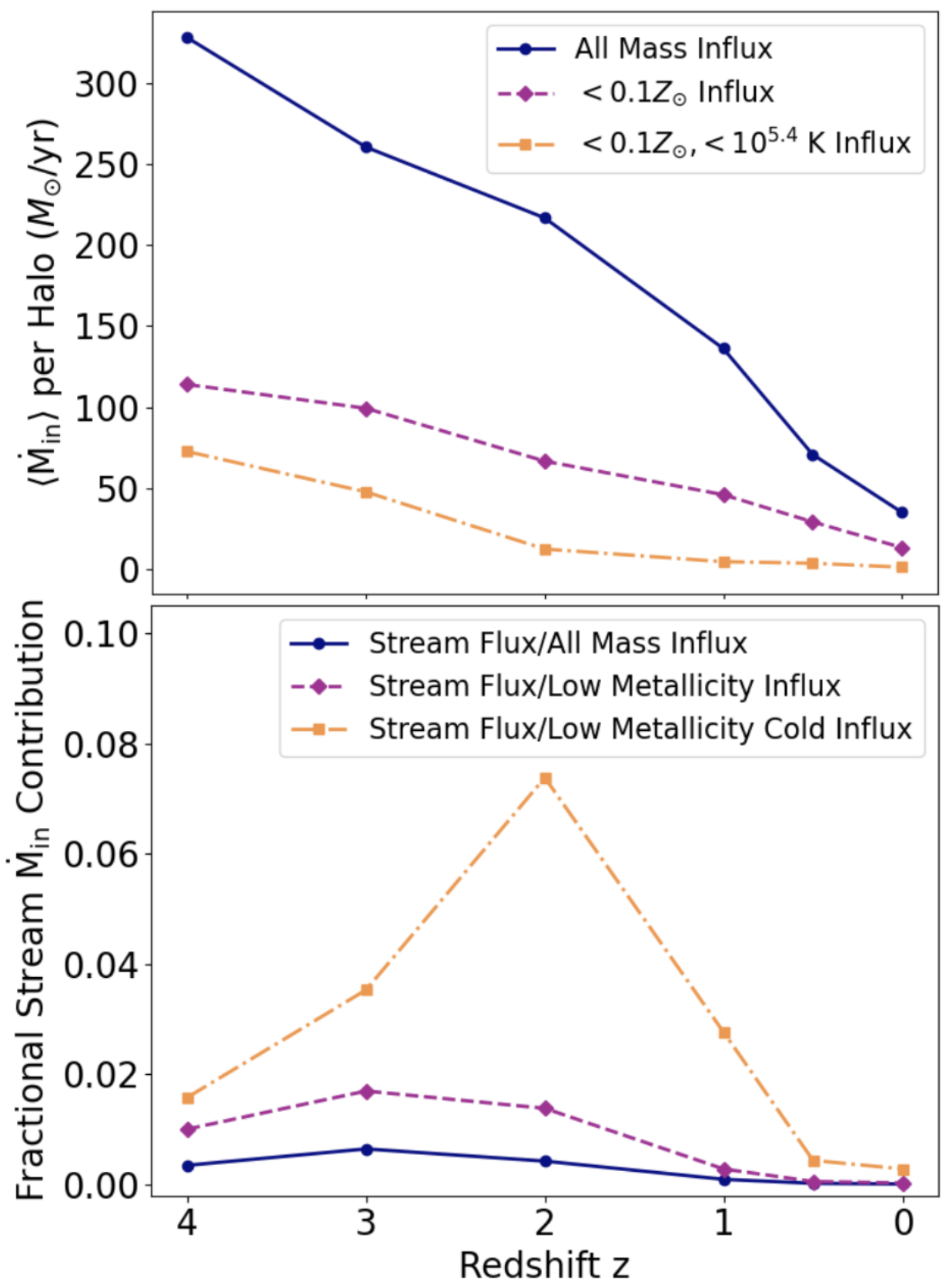

**Figure 9.** Top panel: mean mass flux at $R_{200c}$ over all halos over cosmic time for all mass influx; low-metallicity ($Z < 0.1Z_\odot$) mass influx; and low-metallicity, cold ($T < 10^{5.4}$ K) mass influx. Bottom panel: the fractional contribution of mass influx of streams to the three mass flux rates presented above.

Looking at the mean accretion rate of streams over time, we see that the stream strength peaks at $z = 3{-}2$, after which it declines significantly. The total mass flux from streams shown in the bottom panel of Figure 10 peaks at $z = 2$, which makes sense, as we have a large quantity of halos with streams, and the stream strength is, on average, higher.

### 4.5. Cold Stream Configuration Geometry

In addition to analyzing the physical properties of the streams, we examine their geometric properties, as shown in Figure 11. We calculate the coplanarity angle of the top three most significant streams (left panel), the radial alignment of streams (middle panel), and the impact parameter of streams, b$\equiv r/R_{200c}$ (right panel).

The left panel of Figure 11 presents the smoothed distributions of the coplanarity angle of the three most significant streams for halos between $z = 4$ and $z = 1$. This angle quantifies the deviation of individual streams from the best-fit plane of the entire stream system. The best-fit plane is found by identifying the primary axis of the three streams and then determining a common plane that minimizes the angular deviation of the stream axes. Small angles (0°–20°) signify a highly coplanar system, as expected in the classical accretion picture, while large angles (>40°) indicate a more isotropic spatial distribution. We perform this calculation only if halos have at least three streams; we exclude data from $z = 0.5$ and $z = 0$ because there is an insufficient number of systems with multiple streams. To determine the three most significant streams, we calculate the mass influx of each stream at $R_{200c}$ and take the three streams with the strongest influx.

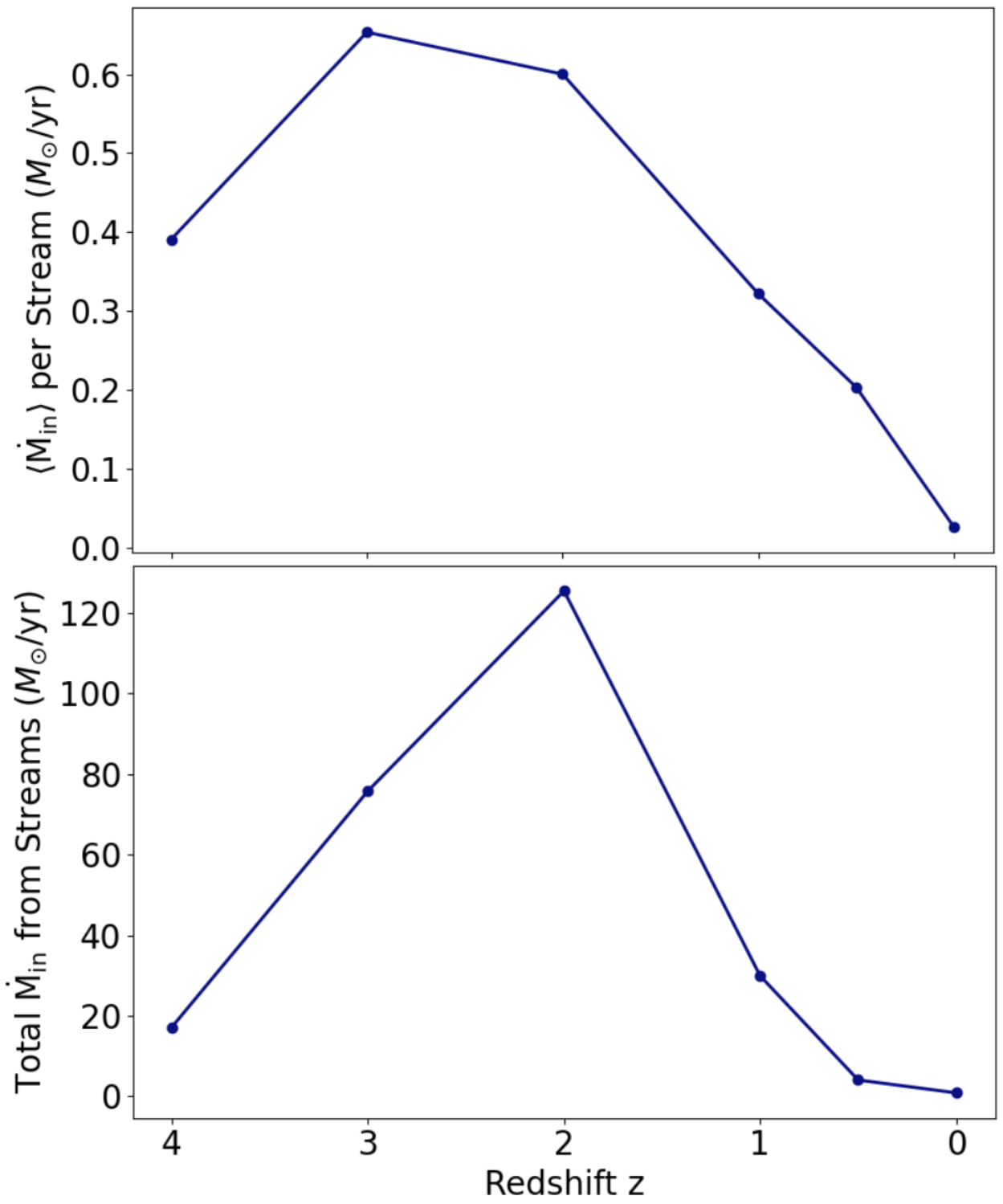

**Figure 10.** Top panel: the mean stream inflow strength at $R_{200c}$ over redshift. Bottom panel: total inflow from streams at $R_{200c}$ in massive halos as a function of redshift. Note that the stream flux is calculated only from cells identified as part of the dense stream core by our algorithm and is therefore a lower limit on the total mass transported by the streams.

At higher redshifts ($z = 3$), stream systems exhibit a relatively uniform distribution in coplanarity up to ~80°. This indicates a mix of different systems, some of which are quite coplanar, while others are more isotropic. Here, it is also important to keep in mind that, at these redshifts, the sample of halos with three or more streams is still rather low. Especially for $z = 4$, there are only three systems with at least three streams to make this measurement, so we are unable to draw significant conclusions. At lower redshifts ($z = 2$–1), there is a more pronounced peak in the distribution toward lower coplanarity angles, especially at $z = 1$. At $z = 2$, the distribution peaks at slightly above ~20°, indicating roughly coplanar stream systems, but the tail extends up to ~80°. At $z = 1$, the distribution peaks slightly below ~20°, again indicating relatively coplanar systems. At this redshift, the tail extends not quite as far, up to ~70°.

The middle panel of Figure 11 shows the smoothed distribution of the radial alignment angle for stream halos at $z = 4$–1. This angle measures how aligned the stream infall direction is with a radial vector from the center of the halo to the stream, defined as

$$\theta_{\rm radial} = \arccos(\hat{r}_{\rm stream} \cdot \hat{r}_{\rm halo\rightarrow stream}), \tag{1}$$

where $\hat{r}_{\rm stream}$ is the stream direction vector, and $\hat{r}_{\rm halo\rightarrow stream}$ is the radial vector from the center of the halo to the stream center of mass. A radial alignment angle of 0° means the streams are perfectly radially infalling into the center of the halo, while 90° indicates purely tangential flow. A small angle (0°–40°) suggests that streams are generally infalling toward the center with some impact parameter. We calculate this for streams in $z = 4$–1, again excluding lower redshifts, due to the small number of streams.

At all redshifts, we find that the radial alignment angle typically peaks at low angles (~15°), indicating that most streams are flowing toward the central galaxy. There is no significant difference between distributions for different redshifts. We would like to note that, as part of the selection process for streams, we discarded streams with very large offsets from the center of the halo. This cut was intended to remove streams that only flow through the outer edges of the halo ($\gtrsim 0.8\,R_{200c}$), and it should not affect the distribution of stream alignment with the halo on scales closer to the center.

Next, we explicitly quantify the impact parameter of these streams, with the smoothed distributions presented in the right panel of Figure 11. We calculate the impact parameter by taking the axis of each stream found through PCA and finding the smallest distance along that axis (assuming that it stretches infinitely) to the halo center. The equation we use to explicitly calculate this is

$$b = ||(r_c - r_{h,\rm center}) \times \hat{r}_{\rm stream}||, \tag{2}$$

where again $\hat{r}_{\rm stream}$ is the normalized stream direction vector, $r_c$ is the centroid point that lies on the stream's axis, and $r_{h,\rm center}$ is the center of mass of the halo, which is 0 in our frame. This distance is then normalized to $R_{200c}$. Once again, we calculate this for streams in $z = 4$–1, excluding lower redshifts, due to the small number of streams.

We find that, for all redshifts, the impact parameter distribution is wide, peaking at roughly b = (0.4–0.5) $r/R_{200c}$ and spanning from very small impact parameters to about $R_{200c}$. It is important to note that this calculation assumes that streams are straight, which is often not the case, as streams can curve. Measuring more accurate impact parameters would require more careful modeling of the trajectories of the streams, including perhaps a velocity flow analysis.

### *4.6. Cold Stream Halo Penetration Efficiency*

If cold streams can indeed supply central galaxies with enough gas to sustain the observed high star formation rates, they must efficiently penetrate deep into halos and deliver gas directly to these galaxies. We perform a detailed analysis of the maximum depth that cold streams reach within halos over cosmic time.

Figure 12 presents the results of this analysis. The left panel shows the mean penetration depth of streams over cosmic time, calculated in three ways: for all detected streams (blue dashed line), for the stream with the highest mass influx in each halo (pink solid line), and for the stream penetrating the deepest within each halo (yellow dotted–dashed line).

We observe that the penetration depth of the streams increases as the redshift decreases to $z = 0.5$, where it peaks, and then decreases slightly to $z = 0$. The significant difference between the "strongest" stream and the "deepest" stream penetration depths indicates that the streams with the highest mass influx are not necessarily the ones penetrating the deepest.

In general, we find that the streams do not penetrate very far into the halos before being disrupted. At $z = 2$, the mean penetration depth of all streams is about $0.8\,R_{200c}$. From the distribution on the right, we see that the deepest any stream penetrates is down to $0.6\,R_{200c}$, while some barely penetrate into $R_{200c}$. Even at $z = 0.5$, where penetration is at its deepest, the mean penetration depth of all streams is around $0.73\,R_{200c}$, while the mean depth for the deepest stream in each halo is about $0.65\,R_{200c}$. We discuss the implications and reasons for this in detail in Section 5.3

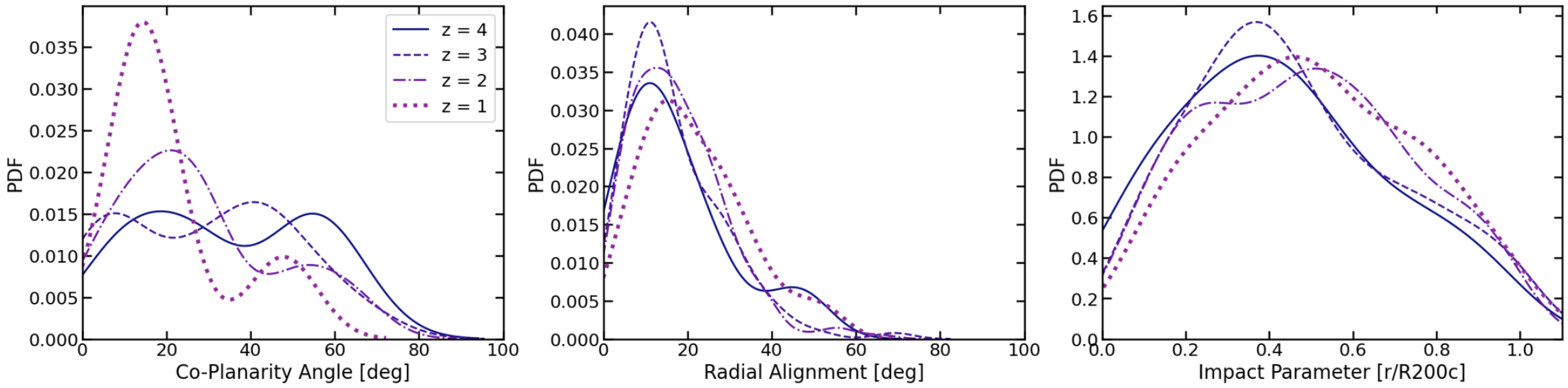

**Figure 11.** Distribution of geometric properties of streams at $z = 4$–1. Left: Coplanarity angle of the three most significant streams per halo. A lower angle indicates a coplanar stream configuration and a higher angle an isotropic configuration. Center: radial alignment angle of streams. Lower angles indicate streams are roughly infalling toward the halo center, and higher angles mean streams are tangential to the halo. Right: impact parameter of streams ($b \equiv r/R_{200c}$), indicating the closest approach of the stream's axis to the center of the halo.

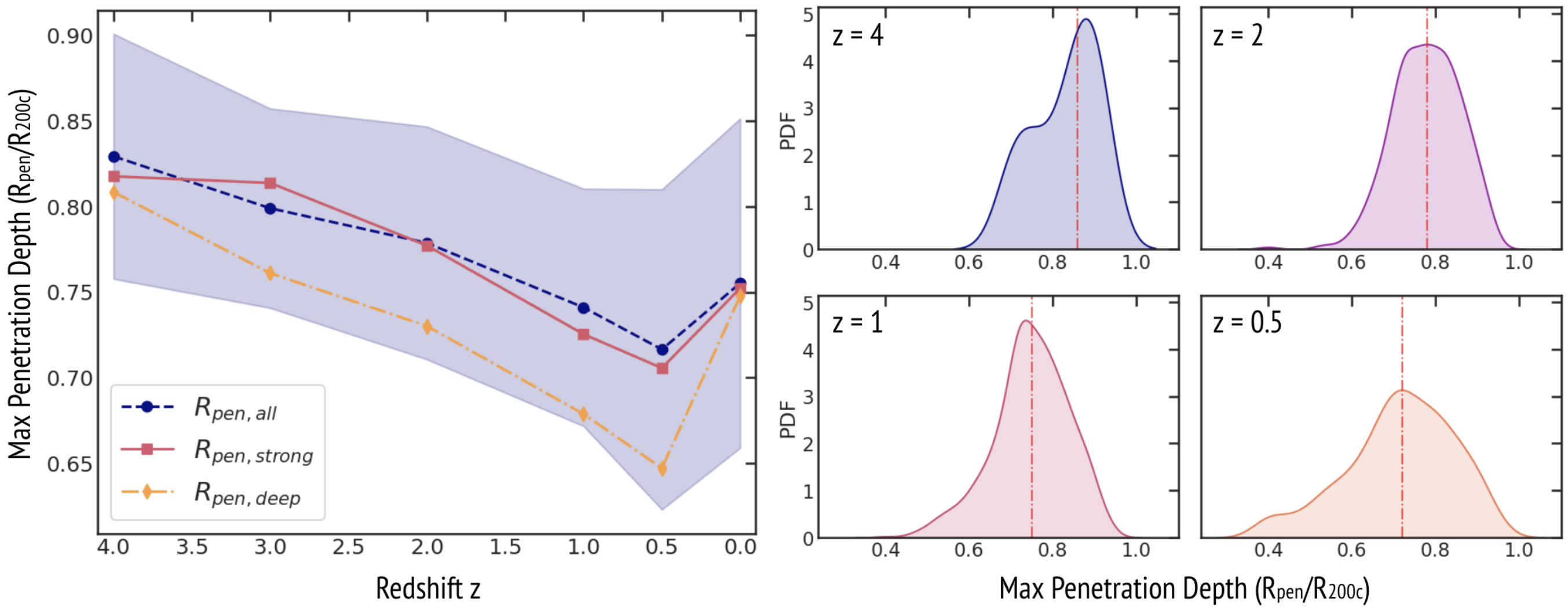

**Figure 12.** Cold stream penetration efficiency. Left: the maximum penetration depth ($R_{\rm pen}/R_{200c}$) of streams over cosmic time, over all streams in each halo (blue dashed line), the "strongest" stream (i.e., the stream with the most mass influx) in each halo (pink solid line), and the deepest-penetrating stream in each halo (yellow dashed–dotted line). The blue shaded region indicates the standard deviation in the mean penetration depth of all streams. Right: distributions of stream penetration at $z = 4$ (top left), $z = 2$ (top right), $z = 1$ (bottom left), and $z = 0.5$ (bottom right).

## 5. Discussion

Having presented our results on the prevalence and properties of cold streams in TNG50, this section places these findings in a broader context. We first discuss the fundamental challenge of defining and detecting accretion streams, then compare our results to previous simulation studies, address the critical caveat of numerical resolution, and conclude by outlining future directions for this work.

### 5.1. The Challenge of Defining and Detecting Streams

The quantitative identification of accretion streams within cosmological simulations is a nontrivial task. The choice of definition—necessarily guided by the scientific question at hand—directly dictates the measured prevalence and physical properties of these structures. The methodology presented in this work is therefore a tailored set of choices developed through extensive testing to navigate the inherent complexities of the problem.

A primary challenge is distinguishing pristine, cosmologically sourced inflows from recycled gas reaccreted from the CGM. As illustrated by Halo 10 (Figure 3), inflowing gas can manifest in distinct morphologies: narrow, filamentary streams and more quasi-spherical, diffuse accretion. A common technique is to apply a metallicity threshold to isolate pristine gas, but this is complicated by turbulent mixing at the stream–CGM boundary, which can entrain enriched material (Z. Hafen et al. 2019; D. Decataldo et al. 2024). This issue is compounded by the difficulty of defining a stream's termination point, as continuous interaction with the CGM leads to thermalization and disruption, blurring the line where a stream ceases to be a coherent structure (D. Padnos et al. 2018; N. Mandelker et al. 2020b; C. Strawn et al. 2021).

Further considerations arise from the morphological diversity and hierarchy of accretion. As shown in Figure 2, streams exhibit a wide range of scales. This diversity necessitates choices that depend on the scientific objective: for example, whether to catalog all discernible filaments or to focus only on those that dominate the mass budget. Similarly, the treatment of merging streams requires a criterion that classifies them as distinct entities or a single larger structure, depending on the scientific focus.

Ultimately, these factors underscore that there is no single, universal definition or detection algorithm for accretion streams. The criteria required to investigate cold, dense streams fueling high-redshift galaxies ($T \sim 10^4$ K) are fundamentally different from those needed to characterize the warm–hot accretion ($T > 10^5$ K) onto massive galaxy clusters. Therefore, adopting a physically motivated and self-consistent framework is paramount to ensure that comparisons of stream properties—across different halos, redshifts, and simulation codes—are both meaningful and consistent.

## 5.2. Comparison with Previous Simulation Studies

### 5.2.1. Stream Prevalence and Configurations

Our finding that the number of streams per halo peaks at $z = 1-2$ contrasts with the work of R. Cen (2014), who found a monotonic decrease in stream number from $z = 4$ to $z \approx 1$. Our result at $z = 2$ is, however, broadly consistent with the average of three significant streams reported by M. Danovich et al. (2012). As discussed above, such discrepancies often stem from the specific criteria used to define streams; our method is sensitive to distinct morphological structures, whereas others may focus on the total inflow rate, which can be dominated by a few massive streams.

This evolution of stream configurations can be understood within the context of large-scale structure. At high redshift, massive halos ($M_h \gtrsim 10^{12} M_\odot$) are rare peaks at the intersections of cosmic filaments (J. R. Bond et al. 1996; V. Springel et al. 2005). Toward lower redshifts, the connectivity of these halos—that is, the number of filaments feeding them—is expected to decline (S. Codis et al. 2018; D. Galárraga-Espinosa et al. 2024), which aligns with our findings at low redshift. At higher redshifts ($z \sim 4$–3), however, we observe fewer streams than at $z = 2$, in contrast to this expectation. This discrepancy might point to a limitation of our stream finding algorithm at early times, or a discrepancy in the simulation itself. It is also important to note that, at these high redshifts, the sample size is quite small and the mean halo mass is lower as well (see Table 1). While the overall decline in cold accretion corresponds to the decline in the cosmic star formation rate (A. Dekel & Y. Birnboim 2006), our findings pinpoint the key transition epoch for stream-dominated accretion in massive halos to be $z \lesssim 1$.

By $z = 0$, cold streams have almost entirely vanished from our sample. This disappearance is likely driven by two concurrent effects: the decreased prevalence of major cosmic filaments feeding massive halos and the increased efficacy of stellar and AGN feedback, which heats and disrupts inflowing gas (F. van de Voort et al. 2011a, 2011b; D. Nelson et al. 2015; C. A. Correa et al. 2018a). At later redshifts ($z < 0.5$), the characteristic halo mass derived via the Press–Schechter formalism is significantly above our cutoff of $M_h > 10^{11.95} M_\odot$. The vanishing of cold streams does not preclude other accretion modes; indeed, at $z \lesssim 0.5$, shock-heating within massive halos becomes efficient, favoring the formation of warm–hot accretion streams not captured by our cold-gas-focused algorithm (e.g., E. Zinger et al. 2016).

### 5.2.2. Stream Properties and Heating

Our observation that stream temperature increases toward lower redshift is consistent with previous studies, which attribute this heating to two primary mechanisms. First, streams are heated by their interaction with the hot CGM and by feedback-driven outflows (D. Nelson et al. 2015, 2016; S. Waterval et al. 2025). Second, streams can be preheated by accretion shocks within cosmic web filaments before entering the halo's virial radius (Y. Birnboim et al. 2016; H. Aung et al. 2024; Y. S. Lu et al. 2024).

Notably, we find that this systematic temperature increase is accompanied by a stream density that remains relatively constant down to $z \approx 0.5$. This density evolution is broadly consistent with the findings of R. Cen (2014), who reported a similar trend. The combination of rising temperatures and constant density implies a significant increase in the thermal pressure of the streams over cosmic time. Disentangling the physical origins of these trends will require higher-resolution simulations capable of resolving the complex hydrodynamics at the stream–CGM interface.

### 5.2.3. Stream Penetration, Disruption, and Flux

The non-monotonic evolution of stream penetration depth, which peaks at $z = 0.5$, presents a particularly interesting puzzle. This peak coincides with both a decline in the mean stream density and a sharp drop in stream prevalence, with only ~50% of halos hosting a stream at this epoch.

We hypothesize that this is a selection effect: at lower redshifts, only the most robust streams survive the journey into an increasingly hot and massive with redshift CGM. These surviving streams may possess a higher initial mass flux or a more coherent structure, enabling them to withstand disruption more effectively. Alternatively, with fewer streams per halo, the reduction in disruptive stream–stream interactions may allow a single dominant stream to penetrate deeper.

Our finding that streams typically dissolve by $0.7\,R_{200c}$ is somewhat higher than the $\sim 0.5\,R_{200c}$ found in other AREPO-based high-resolution zoom-in simulations (D. Nelson et al. 2016), and significantly different from the $\sim 0.3\,R_{200c}$ reported from ART simulations (M. Danovich et al. 2012). More recently, S. Waterval et al. (2025) find that streams can penetrate to galaxy scales of $\sim 0.2\,R_{200c}$ in the GASOLINE-based HELLO zoom-in simulations (S. Waterval et al. 2024). This diversity in penetration depths across simulation codes illustrates the points made earlier: the results are highly sensitive to both stream definitions and the numerical methods used to model their underlying physics.

Paired with this lack of efficient halo penetration is an apparent low mass flux being carried from streams in TNG50. This would seem to suggest that cold streams do not play an important role in building disks.

To contextualize our results, we start by comparing them to recent findings from the HELLO and NIHAO zoom-in simulations, both of which use the GASOLINE code (L. Wang et al. 2015; S. Waterval et al. 2024). S. Waterval et al. (2025) find a total mass accretion rate comparable to ours, presented in Figure 8, but in the HELLO simulations, the typical cold gas mass flux is around 50–100 $M_\odot$ yr$^{-1}$ during cosmic noon and drops below 10 $M_\odot$ yr$^{-1}$ at $z = 0$. They do not explicitly calculate the rate from cold streams, but conclude that most of this cold accretion is in the form of cold streams at high $z$. This discrepancy is likely due to the higher resolution in the HELLO simulations, paired with the different cooling prescription, which allows for a higher cold gas fraction.

When making this comparison, and in general, interpreting our results, there are three important points to consider. First, as discussed previously, we calculate the stream flux by explicitly adding up cells that have been identified as belonging to streams by our algorithm, which underestimates the flux. Second, we must consider the resolution of the simulations, which we will discuss in more detail in the subsequent section. Third, we must consider the cooling prescription in our simulations. In TNG50, there is a cooling floor of $10^4$ K, while in the HELLO simulations, there is a cooling floor of 10 K. Indeed, S. Waterval et al. (2025) find that much of the cold stream gas has a temperature below $10^4$ K. With this in mind, TNG50 is useful for quantifying whether streams appear, but it is not a reliable source for studying stream dynamics and their importance on galactic scales.

### 5.3. Caveats: The Impact of Numerical Resolution

A crucial caveat to our findings is the difficulty of resolving the multi-scale physics of cold streams. These structures span from Mpc-long filaments down to kpc-wide channels, yet their fate is governed by hydrodynamical instabilities that develop on scales of $\lesssim$100 pc (N. Mandelker et al. 2016, 2018, 2019a, 2020b, 2020b; H. Aung et al. 2019, 2024). Most large-volume simulations prioritize resolution in dense galactic disks, leaving the critical stream–CGM interaction layers poorly resolved.

The TNG50 simulation exemplifies this limitation. The typical cell size in the outer CGM at the key epoch of $z = 2$–3 is approximately 1 kpc, which is insufficient to capture the onset and growth of the Kelvin–Helmholtz and Rayleigh–Taylor instabilities expected to shred the streams.

State-of-the-art zoom-in simulations are beginning to bridge this resolution gap using techniques like forced refinement (e.g., FOGGIE) or super-Lagrangian methods (e.g., SURGE and GIBLE) (M. S. Peeples et al. 2019; J. Suresh et al. 2019; F. van de Voort et al. 2019; R. Ramesh & D. Nelson 2024). However, even these pioneering efforts do not yet consistently achieve the sub-100 pc resolution throughout the entire CGM needed to fully model stream fragmentation, nor do they focus on lower-mass halos than those studied here. Therefore, while our results provide a valuable large-scale statistical view, the small-scale physics of stream dissolution remains a frontier for future simulations.

### 5.4. Future Directions

To overcome the limitations discussed above, our next step is to conduct a suite of high-resolution zoom-in simulations. We are resimulating 20 halos from our TNG50 sample ($M_h \gtrsim 10^{12} M_\odot$ at $z = 2$), enhancing the mass and spatial resolution by factors of 8 and 2, respectively. Crucially, we will impose a spatial refinement criterion of ∼50 pc throughout the CGM, reaching the resolution of idealized stream studies (N. Mandelker et al. 2016, 2018, 2020b, 2020b; H. Aung et al. 2019, 2024).

We will also enhance the temporal resolution of the simulation snapshots to a cadence of 10–25 Myr. This is essential for accurately capturing the evolution of the stream and its mixing layer with the CGM. Preliminary work on a single halo with a moderate 250 pc CGM resolution already shows an increased stream penetration depth (from 0.8 $R_{200c}$ to 0.5 $R_{200c}$), motivating these more advanced simulations.

Beyond resolution, this future work will also explore the impact of subgrid physics and cooling. Feedback from star formation and AGN plays a critical role in disrupting streams (F. van de Voort et al. 2011a, 2011b; D. Nelson et al. 2015; C. A. Correa et al. 2018a). By running our zoom-in simulations with varied subgrid feedback models, we can better constrain the complex interplay between these internal processes and external accretion, providing a more comprehensive picture of how massive galaxies are fueled.

## 6. Conclusions

Cold, dense streams of gas are predicted to penetrate deeply into massive ($\gtrsim 10^{12} M_\odot$) hot halos, in particular during the peak of cosmic star formation activity (cosmic noon, $z \sim 3$–2), providing the gas necessary to sustain high star formation rates. Using the TNG50 simulations, we investigated the prevalence and characteristics of these cold streams in massive halos across the redshift range $z = 4$–0, encompassing the period around cosmic noon. Our study employed a novel algorithm based on HDBSCAN to automatically detect cold streams in halos. We summarize our key findings below:

1. Cold streams display a wide variety of geometric configurations, including curves, bends, and kinks. Most streams exhibit a colder, denser core surrounded by a warmer, more diffuse envelope (Figures 2, 3, and 4).
2. Cold streams are prevalent in massive hot halos at high redshifts ($z = 4$–1), with their frequency peaking at $z = 1$ (95%). Their prevalence drops significantly at lower redshifts, falling to approximately 50% at $z = 0.5$ and becoming rare at $z = 0$ (Figure 5).
3. The number of streams per halo decreases as redshift decreases. At $z = 2$–1, three-stream configurations are predominant, while at $z = 0$, halos typically host only a single stream. At $z = 4$, two-stream configurations are more common (Figure 5).
4. The streams' average temperature increases with decreasing redshift ($T \simeq 10^4$–$10^{4.8}$ K), while the mean density remains constant from $z = 4$–1 and slightly decreases thereafter (Figure 7). At lower redshifts, there is a dependence on host halo mass, with generally more-massive halos hosting slightly colder and denser streams.
5. The mean inflow strength of streams declines over cosmic time, with the total inflow strength from streams peaking at $z = 2$ (Figure 10).
6. Halos with at least three cold streams exhibit a low coplanarity angle at $z = 2$–1, indicating that the streams are roughly in the same plane. At $z = 4$–3, many streams have higher coplanarity angles, suggesting more isotropic stream configurations (Figure 11).
7. Cold streams penetrate most deeply into halos at $z \sim 0.5$. However, the limited spatial resolution in the outer CGM, along with the relatively high cooling floor in TNG50, constrains our ability to fully resolve small-scale hydrodynamic instabilities and stream–CGM interactions (Figure 12).

Our findings underscore the critical role of cold streams in fueling galaxies at early epochs and the need for higher-resolution simulations to better capture their survival and impact at later times. Future cosmological zoom-in simulations, with targeted CGM refinement, are essential to resolve the turbulent mixing layers and feedback–inflow interactions that determine whether cold streams can sustain star formation by reaching the galactic disk.

## Acknowledgments

We acknowledge the support of the NSF grant AST-2307280 and the facilities and staff of the Yale Center for Research Computing (YCRC). Some of the results in this paper have been derived using the healpy and HEALPix packages.

## Appendix
## Distinguishing Stream Gas from Recycled Gas

One key challenge in identifying cold gas streams is distinguishing between pristine inflow from the intergalactic medium (IGM) and recycling-enriched gas originating from within the halo. Halo 10, depicted in the top row of Figure 3, serves as a good example. The middle panel of the mass flux projection map (with mass inflow in blue) reveals thin stream structures, along with more isotropic inflow near the halo center. To test our ability to distinguish between recycled and stream gas, we show the same map with radial velocity and metallicity cuts in Figure 13. The left panel includes all the gas in the box. In the remaining two panels, we first apply a radial velocity cut to isolate only the inflowing gas. Then, we apply two different cuts on metallicity: $<0.5\,Z_{\odot}$ (middle panel) and $<0.05\,Z_{\odot}$ (right panel). These figures show that these cuts effectively remove the recycled gas, especially with the more stringent cut, while preserving the inflowing streams. Please note that these cuts are for illustrative purposes and not used in the final stream identification algorithm.

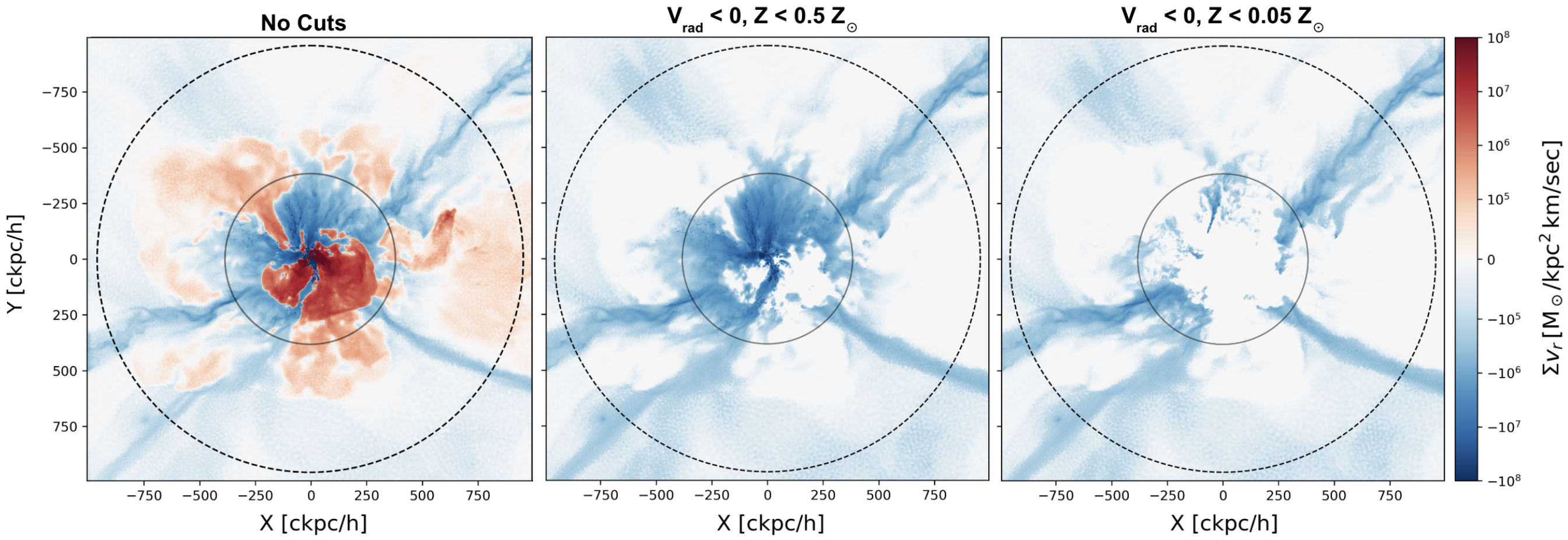

**Figure 13.** Test of metallicity cuts to distinguish stream gas from recycled gas in Halo 10 from the top row of Figure 3. The mass flux map has a projection depth of $0.1\,R_{200c}$, extending $2.6\,R_{200c}$ on each side. The left map displays all gas. The middle and right maps isolate only inflowing gas with metallicity below $0.5\,Z_{\odot}$ (middle) and $0.05\,Z_{\odot}$ (right).

## ORCID iDs

Isabel Medlock https://orcid.org/0009-0000-7415-8239
Daisuke Nagai https://orcid.org/0000-0002-6766-5942
Nir Mandelker https://orcid.org/0000-0001-8057-5880
Volker Springel https://orcid.org/0000-0001-5976-4599
Elad Zinger https://orcid.org/0000-0002-6316-3996
Barry T. Chiang https://orcid.org/0000-0002-9370-4490